\documentclass[12pt]{iopart}
\usepackage{amsthm}
\usepackage{amssymb}
\usepackage{mathtools}
\usepackage{graphicx}
\usepackage[left=25mm,right=25mm,top=25mm,columnsep=15pt]{geometry} 
\usepackage{adjustbox}
\usepackage{placeins}
\usepackage[T1]{fontenc}
\usepackage{lipsum}
\usepackage{csquotes}
\usepackage{hyperref}
\hypersetup{
    colorlinks=true,
    allcolors=blue,
    }
\usepackage[subrefformat=parens,labelformat=parens,caption=false]{subfig}
\usepackage[square,sort&compress,comma,numbers]{natbib}

\begin{document}
\title[Dissipation on a driven qubit: assessment of weak-coupling approximations]{Assessment of weak-coupling approximations on a driven two-level system under dissipation}

\author{W S Teixeira$^{1,2}$, F L Semi\~ao$^1$, J Tuorila$^{2,3}$, M M\"ott\"onen$^{2,4}$}

\address{$^1$ Centro de Ci\^encias Naturais e Humanas, Universidade Federal do ABC, Santo Andr\'e, 09210-170 S\~ao Paulo, Brazil}
\address{$^2$ QCD Labs, QTF Centre of Excellence, Department of Applied Physics, Aalto University, P.O. Box 15100, FI-00076 Aalto, Finland}
\address{$^3$ IQM, Keilaranta 19, FI-02150 Espoo, Finland}
\address{$^4$ VTT Technical Research Centre of Finland Ltd., QTF Center of Excellence, P.O. Box 1000, FI-02044 VTT, Finland}
\ead{wallace.santosteixeira@aalto.fi}
	
	\begin{abstract}
		The standard weak-coupling approximations associated to open quantum systems have been extensively used in the description of a two-level quantum system, qubit, subjected to relatively weak dissipation compared with the qubit frequency. However, recent progress in the experimental implementations of controlled quantum systems with increased levels of on-demand engineered dissipation has motivated precision studies in parameter regimes that question the validity of the approximations, especially in the presence of time-dependent drive fields. In this paper, we address the precision of weak-coupling approximations by studying a driven qubit through the numerically exact and non-perturbative method known as the stochastic Liouville-von Neumann equation with dissipation. By considering weak drive fields and a cold Ohmic environment with a high cutoff frequency, we use the Markovian Lindblad master equation as a point of comparison for the SLED method and study the influence of the bath-induced energy shift on the qubit dynamics. We also propose a metric that may be used in experiments to map the regime of validity of the Lindblad equation in predicting the steady state of the driven qubit. In addition, we study signatures of the well-known Mollow triplet and observe its meltdown owing to dissipation in an experimentally feasible parameter regime of circuit electrodynamics. Besides shedding light on the practical limitations of the Lindblad equation, we expect our results to inspire future experimental research on engineered open quantum systems, the accurate modeling of which may benefit from non-perturbative methods.
	\end{abstract}
	
	

\global\long\def\ket#1{|#1\rangle}%

\global\long\def\Ket#1{\left|#1\right>}%

\global\long\def\bra#1{\langle#1|}%

\global\long\def\Bra#1{\left<#1\right|}%

\global\long\def\bk#1#2{\langle#1|#2\rangle}%

\global\long\def\BK#1#2{\left\langle #1\middle|#2\right\rangle }%

\global\long\def\kb#1#2{\ket{#1}\!\bra{#2}}%

\global\long\def\KB#1#2{\Ket{#1}\!\Bra{#2}}%

\global\long\def\mel#1#2#3{\bra{#1}#2\ket{#3}}%

\global\long\def\MEL#1#2#3{\Bra{#1}#2\Ket{#3}}%

\global\long\def\n#1{|#1|}%

\global\long\def\N#1{\left|#1\right|}%

\global\long\def\ns#1{|#1|^{2}}%

\global\long\def\NS#1{\left|#1\right|^{2}}%

\global\long\def\nn#1{\lVert#1\rVert}%

\global\long\def\NN#1{\left\lVert #1\right\rVert }%

\global\long\def\nns#1{\lVert#1\rVert^{2}}%

\global\long\def\NNS#1{\left\lVert #1\right\rVert ^{2}}%

\global\long\def\ev#1{\langle#1\rangle}%

\global\long\def\EV#1{\left\langle #1\right\rangle }%

	\global\long\def\ha{\hat{a}}%

\global\long\def\hb{\hat{b}}%

\global\long\def\hc{\hat{c}}%

\global\long\def\hd{\hat{d}}%

\global\long\def\he{\hat{e}}%

\global\long\def\hf{\hat{f}}%

\global\long\def\hg{\hat{g}}%

\global\long\def\hh{\hat{h}}%

\global\long\def\hi{\hat{i}}%

\global\long\def\hj{\hat{j}}%

\global\long\def\hk{\hat{k}}%

\global\long\def\hl{\hat{l}}%

\global\long\def\hm{\hat{m}}%

\global\long\def\hn{\hat{n}}%

\global\long\def\ho{\hat{o}}%

\global\long\def\hp{\hat{p}}%

\global\long\def\hq{\hat{q}}%

\global\long\def\hr{\hat{r}}%

\global\long\def\hs{\hat{s}}%

\global\long\def\hu{\hat{u}}%

\global\long\def\hv{\hat{v}}%

\global\long\def\hw{\hat{w}}%

\global\long\def\hx{\hat{x}}%

\global\long\def\hy{\hat{y}}%

\global\long\def\hz{\hat{z}}%

\global\long\def\hA{\hat{A}}%

\global\long\def\hB{\hat{B}}%

\global\long\def\hC{\hat{C}}%

\global\long\def\hD{\hat{D}}%

\global\long\def\hE{\hat{E}}%

\global\long\def\hF{\hat{F}}%

\global\long\def\hG{\hat{G}}%

\global\long\def\hH{\hat{H}}%

\global\long\def\hI{\hat{I}}%

\global\long\def\hJ{\hat{J}}%

\global\long\def\hK{\hat{K}}%

\global\long\def\hL{\hat{L}}%

\global\long\def\hM{\hat{M}}%

\global\long\def\hN{\hat{N}}%

\global\long\def\hO{\hat{O}}%

\global\long\def\hP{\hat{P}}%

\global\long\def\hQ{\hat{Q}}%

\global\long\def\hR{\hat{R}}%

\global\long\def\hS{\hat{S}}%

\global\long\def\hT{\hat{T}}%

\global\long\def\hU{\hat{U}}%

\global\long\def\hV{\hat{V}}%

\global\long\def\hW{\hat{W}}%

\global\long\def\hX{\hat{X}}%

\global\long\def\hY{\hat{Y}}%

\global\long\def\hZ{\hat{Z}}%

\global\long\def\hap{\hat{\alpha}}%

\global\long\def\hbt{\hat{\beta}}%

\global\long\def\hgm{\hat{\gamma}}%

\global\long\def\hGm{\hat{\Gamma}}%

\global\long\def\hdt{\hat{\delta}}%

\global\long\def\hDt{\hat{\Delta}}%

\global\long\def\hep{\hat{\epsilon}}%

\global\long\def\hvep{\hat{\varepsilon}}%

\global\long\def\hzt{\hat{\zeta}}%

\global\long\def\het{\hat{\eta}}%

\global\long\def\hth{\hat{\theta}}%

\global\long\def\hvth{\hat{\vartheta}}%

\global\long\def\hTh{\hat{\Theta}}%

\global\long\def\hio{\hat{\iota}}%

\global\long\def\hkp{\hat{\kappa}}%

\global\long\def\hld{\hat{\lambda}}%

\global\long\def\hLd{\hat{\Lambda}}%

\global\long\def\hmu{\hat{\mu}}%

\global\long\def\hnu{\hat{\nu}}%

\global\long\def\hxi{\hat{\xi}}%

\global\long\def\hXi{\hat{\Xi}}%

\global\long\def\hpi{\hat{\pi}}%

\global\long\def\hPi{\hat{\Pi}}%

\global\long\def\hrh{\hat{\rho}}%

\global\long\def\hvrh{\hat{\varrho}}%

\global\long\def\hsg{\hat{\sigma}}%

\global\long\def\hSg{\hat{\Sigma}}%

\global\long\def\hta{\hat{\tau}}%

\global\long\def\hup{\hat{\upsilon}}%

\global\long\def\hUp{\hat{\Upsilon}}%

\global\long\def\hph{\hat{\phi}}%

\global\long\def\hvph{\hat{\varphi}}%

\global\long\def\hPh{\hat{\Phi}}%

\global\long\def\hch{\hat{\chi}}%

\global\long\def\hps{\hat{\psi}}%

\global\long\def\hPs{\hat{\Psi}}%

\global\long\def\hom{\hat{\omega}}%

\global\long\def\hOm{\hat{\Omega}}%

\global\long\def\hdgg#1{\hat{#1}^{\dagger}}%

\global\long\def\cjg#1{#1^{*}}%

\global\long\def\hsgx{\hat{\sigma}_{x}}%

\global\long\def\hsgy{\hat{\sigma}_{y}}%

\global\long\def\hsgz{\hat{\sigma}_{z}}%

\global\long\def\hsgp{\hat{\sigma}_{+}}%

\global\long\def\hsgm{\hat{\sigma}_{-}}%

\global\long\def\hsgpm{\hat{\sigma}_{\pm}}%

\global\long\def\hsgmp{\hat{\sigma}_{\mp}}%

\global\long\def\dert#1{\frac{\textrm{d}}{\textrm{d}t}#1}%

\global\long\def\dertt#1{\frac{\textrm{d}#1}{\textrm{d}t}}%

\global\long\def\Tr{\text{Tr}}%

	\maketitle
	
	\section{Introduction} \label{sec:intro}
Driven quantum systems are ubiquitous in quantum technologies. They appear, for example, in the control and measurement protocols as well as in the studies of non-equilibrium dynamics~\cite{CrispinGardiner2015,Weiss2008}. One of the simplest paradigmatic examples encompasses a two-level quantum system, a qubit, subjected to a classical drive field which promotes population dynamics in the eigenbasis of the bare qubit. Despite its simplicity, such a model has been applied in many contexts ranging from the coherent control in quantum computing to the simulation of a number of important photochemical reactions~\cite{Cirac1995,Blais2004,Nakamura1999,Cole2001,Letchumanan2004,Eckel2009,Medina2019,Gelman2005,Prokhorenko2006,Golubev2015}. 
Moreover, its properties have been investigated through different descriptions such as the dressed and Floquet state formalisms~\cite{Wilson2010,Grifoni1998,Silveri2017,Nakamura2001,Tuorila2010,Deng2015}, also being associated with various physical phenomena, such as coherent suppression of tunneling~\cite{Grossmann1991,Dakhnovskii1993,Grifoni1998} and interference between successive Landau--Zener transitions~\cite{Shevchenko2010,Oliver2005,Sillanpaa2006,Silveri2015,Tuorila2013}.

Another well-known example of drive-induced quantum phenomena is attributed to the work by Mollow in Ref.~\cite{Mollow1969}.
Remarkably, Mollow theoretically showed that the fluorescence spectrum of a driven qubit may turn into a triplet in the presence of weak dissipation. If the Rabi frequency of the classical field well exceeds the dissipation rate, two sidebands emerge in the spectrum with an offset equal to the Rabi frequency from the center peak at the drive frequency. A more sophisticated explanation of such a phenomenon was later provided using a quantum treatment also for the drive field~\cite{Oliver1971,Carmichael1976,Cohen-Tannoudji1977}. In this so-called dressed-state picture, the energy levels of the composite qubit-field system are split due to the dynamic Stark effect promoted by the drive. The Mollow triplet has been verified experimentally in many different physical scenarios~\cite{Schuda1974,Wrigge2007,Baur2009,Astafiev2010,Ulhaq2013,Unsleber2015,Pigeau2015,Lagoudakis2017,Ortiz-Gutierrez2019}.

The approach for solving the open-quantum-system dynamics in Mollow's study~\cite{Mollow1969} and in the follow-up work in Refs.~\cite{Oliver1971,Carmichael1976,Cohen-Tannoudji1977} assumes a weak coupling between the system, i.e., the qubit, and its bath of quantized bosonic modes, thus motivating a perturbative treatment of the dissipation~\cite{Carmichael2013,Heinz-PeterBreuer2007,CrispinGardiner2004}. Such an approach is guided by the so-called Born--Markov approximations, where one assumes a stationary bath and fast decay of bath correlations in the typical timescales of the system evolution. Furthermore, upon the elimination of fast-oscillating terms, the typically non-unitary evolution of the system is usually expressed by a Lindblad master equation (LME)~\cite{Lindblad1976,Gorini1976} with positive decay, excitation, and dephasing rates, in addition to which the environment introduces a rescaling of the transition frequency of the free system.

However, one of the possible drawbacks of such a form of the LME is that the interplay between the drive and dissipation is not fully contemplated. As a consequence, the presence of a time-dependent drive field may rise questions on the validity of the above-mentioned approximations. In the literature, a vast amount of strategies have been presented to approach such a scenario, each with their own range of applicability and assumptions motivated by the details of the system under study.
For example, still within the Born--Markov approximations, the effects of a strong drive on the open dynamics of a superconducting qubit has been investigated in the dressed-state picture~\cite{Wilson2010,Nakamura2001}. Here, the environmental effects on the dressed qubit-field states become equivalent to the case of a weakly dissipative and non-driven qubit if the field has a sufficiently large average photon number~\cite{Wilson2010}. The Born--Markov approximations have also been a starting point of other approaches for the open dynamics of slowly~\cite{Pekola2010,Salmilehto2010,Salmilehto2011,Albash2012,Xu2014} and periodically driven quantum systems~\cite{Blumel1991,Dittrich1993,Grifoni1998,Hausinger2010,Engelhardt2019}, the latter with dissipative effects manifested through incoherent transitions between the Floquet states of the time-dependent system Hamiltonian. Analytical developments have also been obtained in the high-driving-frequency regime~\cite{Hartmann2000, Ikeda2020}. Despite the great efforts to treat dissipation in driven systems, to some extent these techniques are perturbative in the system-bath coupling, which does not allow for a precise benchmark of the weak-coupling assumptions of the open quantum dynamics. 

Perturbative expansions of system-bath couplings in the presence of time-dependent fields may be overcome through the noninteracting-blip approximation (NIBA)~\cite{Leggett1987,Leggett1995,Grifoni1995,Grifoni1998,Magazzu2018b} or by suitable changes of frame of reference, including polaron-type transformations~\cite{McCutcheon2011,McCutcheon2013}, fast drive rotations~\cite{Restrepo2016}, and exact mappings of the bath coordinates onto one-dimensional chains~\cite{Prior2010,Chin2010}. Furthermore, the dynamics of the dissipative driven qubit outside the Born--Markov approximations has also been extensively studied through other numerical methods such as the quasiadiabatic propagator path integral (QUAPI)~\cite{Makarov1994,Makri1995,Makri1995b,Grifoni1998}, the hierarchical equations of motion (HEOM)~\cite{Tanimura1989,Tanimura2020}, and through a direct discretization of the bath modes~\cite{Cangemi2019}. Numerical methods based on matrix product operator techniques have also been recently proposed~\cite{Strathearn2018,Cygorek2021}.

Given the wide range of available techniques for the study of driven dissipative quantum systems, in this paper we use the well-established stochastic Liouville equation with dissipation (SLED)~\cite{Stockburger1999} to investigate the dynamics and steady-state properties of a dissipative driven qubit. Assuming a linear system--bath interaction and factorized initial states, such a method has the advantage of being non-perturbative and numerically exact provided that the spectral density of the bath is Ohmic with a high cutoff frequency. Within SLED, time-dependent drive fields can be included without further assumptions on its parameters. The SLED and related methods have been used, for instance, in the study of tunneling~\cite{Stockburger1999b} and of the optimal control of quantum systems~\cite{Schmidt2011,Schmidt2013}. 

More recently, the SLED has also been employed in benchmarks for the initialization of a non-driven superconducting qubit~\cite{Tuorila2019} and in the validity check of weak-coupling approaches for the open dynamics of a single and two non-driven qubits~\cite{Vadimov2021}. In this paper, we focus on the case where a single qubit is weakly driven by nearly resonant transverse fields and the interaction with a cold Ohmic bath produces effective dissipation rates that reach up to $10\%$ of the bare qubit angular frequency. Experimentally, this scenario has been motivated by the recent progress in the implementation of tunable and engineered environments, for example, in circuit quantum electrodynamics (cQED)~\cite{Viitanen2021,Partanen2019,Sevriuk2019,Silveri2019,Partanen2018,Tan2017,Harrington2019,Martinez2019,Ronzani2018,Kimchi-Schwartz2016,Murch2012}. Note that another numerically exact and non-perturbative method has been proposed to capture more general initial system-bath states, including correlated ones~\cite{Orth2013}. In this paper, however, we restrict our studies to the case of factorized initial states in such a way that the use of SLED is well justified. In principle, one has the choice to prepare such a factorized state in the beginning of the dynamics.

In this context, the goal of this work is the following: Firstly, we investigate the main characteristics introduced by SLED on the properties of the driven qubit by considering the usual LME~\cite{Carmichael2013,Heinz-PeterBreuer2007} as a point of comparison. To this end, we show that the often overlooked bath-induced energy shift term in the LME may give an important contribution to the dynamics and greatly alter the actual steady state of the driven qubit, in stark contrast to non-driven systems with otherwise matching parameter values. Even though the relevance of the bath-induced energy shift has already been pointed out in other contexts, e.g.~\cite{Vega2010,Thingna2012}, our study illustrates the regimes where such effects are more pronounced, implying that they should be carefully considered in the related experiments. Taking advantage of the non-perturbative characteristics of SLED, we also propose a scheme to experimentally witness the failure of the asymptotic predictions of the LME. Secondly, we study signatures of the Mollow triplet using the SLED in a pump-probe spectroscopy configuration with parameter regimes that may be implemented in the framework of cQED. 

This manuscript is organized as follows. In section~\ref{sec:model}, we present the theoretical and numerical models, emphasizing the main differences between the Lindblad and SLED master equations for a driven system. In section~\ref{sec:monofield}, we apply both methods in the case of a monochromatic transverse drive field and show that the overlap fidelity between the SLED and Lindblad solutions is drastically reduced if the bath-induced energy shift term is not taken into account in the LME.
Furthermore, we analyse the steady-state properties of the driven system and describe a protocol to witness the failure of the LME. The concern here is not to prove that the approximate treatment with the LME fails, but instead to show the practical relevance of non-perturbative approaches such as the one we utilize in this work. In section~\ref{sec:mollow}, we numerically study the meltdown of the Mollow triplet within the SLED and Lindblad formalisms for experimentally feasible parameters for cQED. Section~\ref{sec:conc} concludes this work.

	\section{Model} \label{sec:model}

Consider a qubit with the bare transition frequency $\omega_{\text{q}}$ driven by a time-dependent transverse field with Hamiltonian $\hH_{\text{d}}(t)$. We express the free Hamiltonian of the qubit with the help of its eigenbasis $\{\ket{0},\ket{1}\}$ as $\hH_\text{S}=-\hbar\omega_{\text{q}}\hsgz/2$, where $\hsgz=\kb{0}{0}-\kb{1}{1}$, and the drive Hamiltonian as $\hH_{\text{d}}(t)=\hbar f(t)\hsgx$, where 
$f(t)$ is a time-dependent function with the units of angular frequency and $\hsgx=\kb{0}{1}+\kb{1}{0}$.

Some particular choices of $\hH_{\text{d}}(t)$ have been historically used in the study of driven quantum systems and in the discovery of novel physical phenomena. For example, with $f(t)=\Omega_0+\Omega_{\text{d}}\cos(\omega_{\text{d}} t)$, the dynamics of the expectation value of $\hsgx$ may be frozen in the fast drive regime $\omega_{\text{d}}\gg \omega_{\text{q}}$ for selected values of $\Omega_{\text{d}}$ and $\omega_{\text{d}}$ even when $\Omega_0\neq0$. This phenomenon is usually referred to as coherent destruction of tunneling~\cite{Grifoni1998}. For $f(t)\propto t$, the transition probability between the lower eigenstate of the instantaneous Hamiltonian $\hH_{\text{S}}+\hH_{\text{d}}(t)$ at $t\rightarrow-\infty$ and its excited eigenstate at $t\rightarrow + \infty$ can be found analytically, a case generally referred to as Landau--Zener transitions~\cite{Shevchenko2010}.
In contexts where the drive field couples weakly to the qubit, such as in laser--atom interactions, and its frequency lies near the bare qubit frequency $\omega_{\text{q}}$, one may apply the rotating-wave approximation and write $\hH_{\text{d}}(t)\approx\hbar\Omega_{\text{d}}(\kb{0}{1}\text{e}^{\text{i}\omega_{\text{d}}t}+\kb{1}{0}\text{e}^{-\text{i}\omega_{\text{d}}t})/2$, as in the original investigation of the qubit fluorescence spectrum in Mollow's work~\cite{Mollow1969}. 

For the background theory considered in this section, we do not consider any specific form of $f(t)$. In sections~\ref{sec:monofield} and~\ref{sec:mollow}, we consider the case of oscillating transverse fields without a constant term. The only assumption made here is that such a classical field is a good approximation of a coherent quantum field over the time scales of interest~\cite{Dutra1994,Salmilehto2014,Ikonen2017}. This restricts the subsequent analysis to a system described by a two-dimensional Hilbert space, thus reducing computational time of the numerically exact protocol.

We choose the qubit to interact linearly with a dissipative bosonic bath which is modeled by an infinite set of quantum harmonic oscillators. The $j$:th oscillator has creation and annihilation operators $\hdgg{b}_j$ and $\hb_j$, respectively, and an angular frequency $\omega_j$. The Hamiltonian of the bare bath can thus be written as $\hH_{\text{B}}=\hbar\sum_{j}\omega_{j}\hdgg b_{j}\hb_{j}$, and the system-bath interaction Hamiltonian as $\hH_{\text{SB}}=\hbar\hsgx\sum_{j}g_{j}(\hb_{j}+\hdgg b_{j})$, with $\{g_j\}$ being the corresponding coupling strengths. The total Hamiltonian is given by
\begin{align}
	\hH(t) & =\hH_{\text{S}}+\hH_{\text{d}}(t)+\hH_{\text{B}}+\hH_{\text{SB}}.\label{eq:H1}
\end{align}

\begin{figure}
	\centering
	\includegraphics[width=0.4\linewidth]{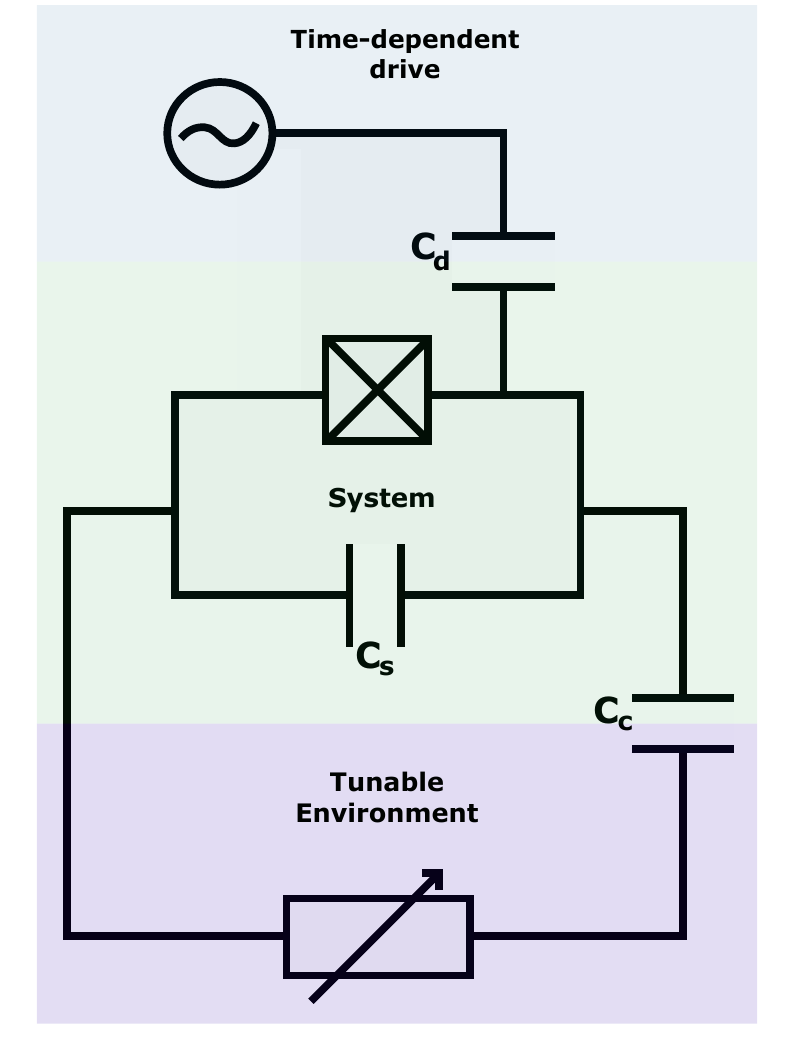}
	\caption{Possible quantum-electric-circuit implementation corresponding to the model considered in this work. A transmon qubit (system) composed of a Josephson junction (boxed cross) and a shunt capacitance $C_{\textrm{s}}$ is coupled to both, an ac voltage source and an effective tunable resistor (environment) through the capacitances $C_{\textrm{d}}$ and $C_{\textrm{c}}$, respectively. Typical parameters for this type of system can be found in table~\ref{tb:pmt}.} 
	\label{fig:system}
\end{figure}

Conveniently, the system-bath interaction can be fully characterized
by the spectral density function 
\begin{align}
	J(\omega)=2\pi\sum_{j}g_{j}^{2}\delta\left(\omega-\omega_{j}\right), \label{eq:J1}
\end{align}
where $\delta(\omega-\omega_{j})$ is Dirac delta function. In the continuum limit, the spectral density becomes a smooth function of $\omega$ that approaches zero as $\omega\rightarrow \infty$. This so-called ultraviolet cutoff is physically motivated in cQED, for example, by the finite bandwidth of the transmission lines coupled to the qubit. Specifically, we use 
an Ohmic environment with a quartic Drude cutoff characterized by the spectral density~\cite{Tuorila2019,Vadimov2021}
\begin{align}
	J(\omega)=\frac{2\eta \omega}{\left(1+\frac{\omega^2}{\omega_\textrm{c}^2}\right)^2}, \label{eq:J2} 
\end{align}
where $\eta$ is an effective dimensionless coupling constant and $\omega_{\textrm{c}}$ is the bath cutoff frequency, which is assumed to be much higher than the qubit bare frequency. The quartic cutoff has been chosen over the typical quadratic Drude and exponential cutoffs in order to speed up the convergence of the numerical method since it produces a conveniently narrow spectrum for a given cutoff frequency $\omega_{\textrm{c}}$. Nevertheless, this spectral density can capture the relevant physics of a tunable resistor coupled to a superconducting transmon qubit \cite{Tuorila2019} shown in figure~\ref{fig:system}.

In this work, we assume that the total density operator is initially factorized, i.e., $\hrh(0)=\hrh_{\text{S}}(0)\otimes\hrh_{\text{B}}(0)$, with $\hrh_{\text{B}}(0)=\text{e}^{-\beta \hH_{\text{B}}}/\Tr[\text{e}^{-\beta \hH_{\text{B}}}]$ being the Gibbs state of the bath at temperature $T=(k_{\rm B}\beta)^{-1}$ and with mean excitation number $\bar{n}(\omega)=(\text{e}^{\hbar\beta \omega}-1)^{-1}$. Consequently, the internal dynamics of the heat bath can be represented by its autocorrelation function
\begin{align}
	L(\tau)=\langle \hat\xi(\tau)\hat \xi(0)\rangle = \frac{1}{2\pi}\int_{-\infty}^{\infty}\textrm{d}\omega\, \text{e}^{-\text{i}\omega \tau}S(\omega), \label{eq:L1}
\end{align}
where $\hat{\xi}=\sum_jg_j(\hat b_j+\hat b_j^{\dag})$ and $S(\omega)=J(\omega)[\bar{n}(\omega)+1]$ is the power spectrum of the noise. The behavior of $L(\tau)$ also plays an important role in the reduced dynamics of the qubit, with its real and imaginary parts, denoted hereafter by $L_\textrm{r}(\tau)$ and $L_\textrm{i}(\tau)$, respectively.

Motivated by experiments, for example those in nuclear magnetic resonance~\cite{Purcell1946,Bloch1946,Bloembergen1948}, 
a perturbative description of the reduced qubit dynamics under the Hamiltonian in equation~\eqref{eq:H1} and the choice of $\hrh(0)$ has been extensively studied in the weak-coupling case, i.e., for $\n{g_{j}/\omega_{\text{q}}}\ll1$~\cite{Bloch1957,Mollow1969a,Agarwal1973,Carmichael2013,Heinz-PeterBreuer2007}.
Such an approach relies on assuming that the system--bath correlations created dynamically are negligible for the dynamics of the system so that in the reduced master equation of the system, we may use the thermal equilibrium state of the environment, i.e., $\hrh_{\text{B}}(t)\approx\hrh_{\text{B}}(0)$. In addition, memory effects on the reduced dynamics arising from the finite decay time of $L(\tau)$ are usually neglected. These assumptions constitute the Born--Markov approximations, and along with the elimination of quickly oscillating terms, produce a time-local master equation for the reduced density operator $\hrh_{\text{S}}(t)=\Tr_{\text{B}}[\hrh(t)]$ of the qubit which can be cast into a Lindblad form as (see~\ref{sec:Lindblad} for details)
\begin{align}
	\dert{\hrh_{\text{S}}(t)}=&-\frac{\text{i}}{\hbar}[\hH_{\text{S}}+\hH_{\text{s}}+\hH_{\text{d}}(t),\hrh_{\text{S}}(t)]\nonumber\\&+\Gamma(\omega_{\text{q}})D_{01}\left[\hrh_{\text{S}}(t)\right]+\Gamma(-\omega_{\text{q}})D_{10}\left[\hrh_{\text{S}}(t)\right], \label{eq:drho4}
\end{align}
where the superoperators $D_{ij}(\hrh)=\kb{i}{j}\hrh\kb{j}{i}-\{\kb{j}{j},\hrh\}/2$ express incoherent transitions between the eigenstates of $\hH_{\text{S}}$ with positive rates $\Gamma(\pm\omega_{\text{q}})=2\text{Re}[\int_{0}^{\infty}{\textrm{d}\tau\, \text{e}^{\pm \text{i} \omega_{\text{q}}\tau}L(\tau)}]$ and, excluding constant terms, $\hH_{\text{s}}=-\hbar\Delta_{\text{s}}\hsgz/2$ is the bath-induced energy shift of the qubit characterized by the correction to its bare frequency $\omega_{\text{q}}$,
\begin{align}
	\Delta_{\text{s}}&=\Lambda(\omega_{\text{q}})-\Lambda(-\omega_{\text{q}})=2\int_{0}^{\infty}\textrm{d}\tau\, \sin(\omega_{\text{q}}\tau)L_\textrm{r}(\tau), \label{eq:LS1}
\end{align}
with $\Lambda(\pm\omega_{\text{q}})=\text{Im}[\int_{0}^{\infty}{\textrm{d}\tau\, \text{e}^{\pm \text{i} \omega_{\text{q}}\tau}L(\tau)}]$. 
Note that such a correction comprises both the vacuum and thermal contributions of the bath, traditionally referred to as Lamb and Stark shifts, respectively.
For a non-driven system characterized by $\hH_{\text{d}}(t)=0$, equation~\eqref{eq:drho4} describes the thermalization process of the qubit with the heat bath, the superoperator of which commutes with the unitary part of the master equation. Consequently, the specific value of the shift $\Delta_{\text{s}}$ does not affect the steady-state quantities since the coherences in the eigenbasis of $\hH_{\text{S}}$ vanish in the limit $t\rightarrow\infty$. For a driven system in contrast, neglecting the shift may lead to incorrect predictions both dynamically and in the steady state as we show in section~\ref{sec:monofield}. 
The assumption of a stationary thermal bath implies the drive field to contribute only to the unitary part of the master equation (\ref{sec:Lindblad}).

As discussed in the introduction, the limitations imposed by the above-mentioned approximations can be handled through different strategies. Here we use the framework of stochastic Liouville equations, where the non-perturbative treatment of the system-bath coupling relies on a stochastic unraveling of the reduced density operator $\hrh_{\text{S}}(t)$. In this procedure, one resorts to the path integral description of quantum mechanics \cite{Feynman1963} to establish a numerically exact model of the open-quantum-system dynamics with the help of a classical stochastic process~\cite{Stockburger1999,Stockburger2002}.
For the Ohmic spectral density in equation~\eqref{eq:J2}, and in the limit $\omega_\textrm{c}\rightarrow \infty$, a single trajectory for the state of the system, $\hrh'_{\text{S}}(t)$, is given by the so-called stochastic Liouville equation with dissipation, SLED, as \cite{Stockburger1999,Tuorila2019}
\begin{align}
	\dert{\hrh'_{\text{S}}(t)}=&-\frac{\text{i}}{\hbar}[\hH_{\text{S}}+\hH_{\text{d}}(t)-\hbar\xi(t)\hsgx,\hrh'_{\text{S}}(t)] \nonumber \\
	{}&-\frac{\eta}{\hbar\beta}\left[\hsgx,\left[\hsgx,\hrh'_{\text{S}}(t)\right]\right]-\text{i}\frac{\eta\omega_{\text{q}}}{2}\left[\hsgx,\left\{\hsgy,\hrh'_{\text{S}}(t)\right\}\right],
	\label{eq:SLED}
\end{align}
where $\xi(t)$ is a real-valued classical random variable with null mean and autocorrelation
\begin{align}
\mathbb{E}\left[\xi(t)\xi(t')\right]&=\int_{0}^{\infty}\frac{\textrm{d}\omega}{2\pi} J(\omega)[\coth(\hbar\beta \omega/2)-2/(\hbar\beta \omega)]\cos[\omega (t-t')], \label{eq:corrfunc}
\end{align}
with $\mathbb{E}[.]$ denoting the ensemble average over the noise trajectories (see~\ref{sec:SLED} for details). Upon a suitable choice of a real-valued kernel, the quantity $\xi(t)$ can be generated from a delta-correlated Gaussian noise, indeed providing for it a fully stochastic interpretation (see~\ref{subsec:app1}). This term in equation~\eqref{eq:SLED} encodes the quantum fluctuations neglected when one treats the bath classically, as in the high-temperature limit through the Caldeira--Leggett master equation~\cite{Heinz-PeterBreuer2007,Caldeira1981,Stockburger1999}. The actual reduced density operator of the system is obtained as $\hrh_{\text{S}}(t)=\mathbb{E}[\hrh'_{\text{S}}(t)]$ in the limit of infinite trajectories. Note that for a single generation of $\xi(t)$, equation~\eqref{eq:SLED} is deterministic, so that parallelization and usual numerical methods for quantum evolution can be combined to speed up the SLED calculations. 

In order to simplify the comparisons between the SLED and Lindblad approaches in the next sections, we establish a connection between the transition rates in equation~\eqref{eq:drho4} and equation~\eqref{eq:SLED}. This is achieved by writing $\Gamma(\omega_{\text{q}})=\gamma[\bar{n}(\omega_{\text{q}})+1]$ and $\Gamma(-\omega_{\text{q}})=\gamma\bar{n}(\omega_{\text{q}})$, with $\gamma=2\eta\omega_{\text{q}}$ being an effective qubit dissipation rate calculated in the limit $\omega_{\textrm{c}}\gg \omega_{\text{q}}$. Consequently, $\eta=\gamma/(2\omega_{\text{q}})$ in equation~\eqref{eq:SLED}. 

Below, we investigate the features promoted by the numerically exact SLED on the reduced state of the qubit for a nearly resonant drive field $\hH_{\text{d}}(t)$ and for different dissipation rates
which are produced by the tunable environment of the qubit. 
We compare the exact predictions of such a method against the approximate LME. To this end, we define the fidelity of a density operator of an approximate solution $\hrh_1$ against the numerically exact SLED solution $\hrh_2$ for qubits as  \cite{Jozsa1994}
\begin{align}
	\mathcal{F}=\Tr[\hrh_1\hrh_2]+2\sqrt{\det(\hrh_1)\det(\hrh_2)}. \label{eq:fid}
\end{align}
The infidelity, $1-\mathcal{F}$, describes the distance between the two states, and hence provides information on the amount of error introduced by the approximations used in the inexact approaches. 
In particular, we compare the fidelities of the Lindblad solutions obtained with and without the inclusion of the bath-induced energy shift defined in equation~(\ref{eq:LS1}). 
In the following, the latter case is referred to as LME--nES.

	\section{Monochromatic periodic field} \label{sec:monofield}

In this section, we compare the Lindblad and SLED approaches for the dissipative dynamics of a driven qubit. Namely, we solve the Lindblad equation~\eqref{eq:drho4} and compare the results with those of SLED~\eqref{eq:SLED} using a sufficiently large number of noise realizations $\xi(t)$ to obtain the reduced system density operator as an average over individual trajectories. Both LME and SLED are solved in the Liouville space~\cite{Mukamel1995} upon second-order Magnus expansions of the discretized time propagator associated to the corresponding Liouvillians. Particularly in the SLED, each noise realization $\xi(t)$ is generated before the temporal evolution as in the recipe in equation~\eqref{eq:xi2}. Here, we focus on the case in which the qubit is driven by a monochromatic and periodic transverse field at angular frequency $\omega_{\text{d}}$ as
\begin{align}
	f(t)=\Omega_{\text{d}}\cos(\omega_{\text{d}} t),
	\label{eq:f1}
\end{align}
where $\Omega_{\text{d}}$ is referred to as the drive Rabi frequency. The corresponding full microscopic Hamiltonian is given by equation~\eqref{eq:H1}.

\begin{table}
	\caption{Default parameter values for the investigations carried out in this work. The chosen values are typical for experimental realizations of superconducting qubits. The resonator parameters are shown for reference and are only used in the phenomenological description of the measurement setup presented in section~\ref{sec:mollow}.
	}
	\begin{centering}
		\begin{tabular}{l l}
			\hline \hline
			qubit transition frequency $\omega_{\text{q}}/(2\pi)$  & $5.0$ GHz\tabularnewline
			\hline 
			resonator frequency $\omega_{\text{r}}/(2\pi)$ & $7.0$ GHz\tabularnewline
			\hline 
			qubit-resonator coupling strength $g/(2\pi)$ & 100 MHz\tabularnewline
			\hline 
			dispersive shift $\chi/(2\pi)$ & $-5.0$ MHz\tabularnewline
			\hline 
			drive Rabi frequency $\Omega_{\text{d}}/(2\pi)$  & $50.0$ MHz\tabularnewline
			\hline 
			probe Rabi frequency $\Omega_{\text{p}}/(2\pi)$ & $5.0$ MHz\tabularnewline
			\hline 
			measurement drive amplitude $\Omega_{\text{m}}/(2\pi)$ & 250 kHz\tabularnewline
			\hline 
			resonator dissipation rate $\kappa$ & $2\pi\times 250$ kHz\tabularnewline
			\hline 
			qubit dissipation rate $\gamma$ & $2\pi\times 50$ MHz\tabularnewline 
			\hline 
			environment cutoff frequency $\omega_\textrm{c}/(2\pi)$ & 250 GHz\tabularnewline
			\hline 
			bath temperature $T$ & $48$ mK\tabularnewline
			\hline \hline
		\end{tabular}
		\par\end{centering}
	\label{tb:pmt}
\end{table}

\begin{figure*}[t]
	\subfloat{\label{fig:dyna}} 
	\subfloat{\label{fig:dynb}}
	\subfloat{\label{fig:dync}}
	\centering
	\includegraphics[width=\linewidth]{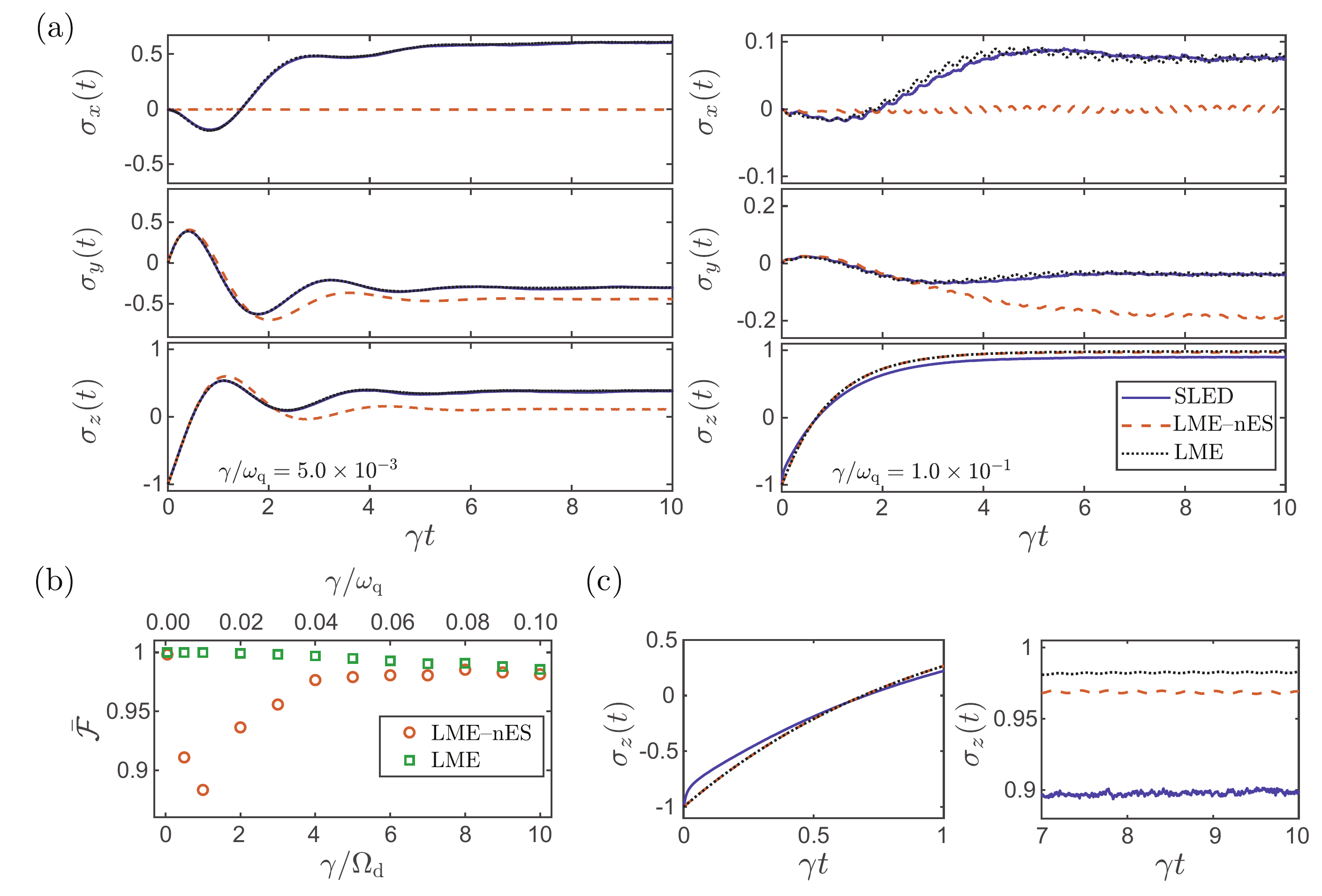}
	\caption{(a) Dynamics of the Bloch vector components in a frame rotating at $\omega_\textrm{d}=\omega_\textrm{q}$ as functions of time $t$ for different indicated values of the dissipation rate $\gamma$. (b) Fidelity of the Lindblad solution against SLED averaged over an interval $t\in[0,10]/\gamma$ 
		as a function of $\gamma$, with (green squares) and without (orangle circles) the bath-induced energy shift. (c) Short-time and long-time dynamics of the $z$ component of the Bloch vector obtained from the right panel of (a). Parameters not given here are chosen as in table~\ref{tb:pmt}.} \label{fig:dyn}
\end{figure*}

In figure~\ref{fig:dyna}, we show the dynamics of the Bloch vector components $\sigma_i(t)=\Tr[\hsg_i\text{e}^{-\text{i}\hsgz\omega_\textrm{d}t/2}\hrh_{\text{S}}(t)\text{e}^{\text{i}\hsgz\omega_\textrm{d}t/2}]$ ($i=x,y,z$ and $\hsgy=\text{i}\kb{1}{0}-\text{i}\kb{0}{1}$) in a frame rotating with the drive frequency $\omega_{\text{d}}$ and for a qubit initially prepared in the excited state $\ket{1}$. 
The drive frequency is set at the bare qubit frequency ($\omega_{\text{d}}=\omega_{\text{q}}$) and the other parameters are chosen in compliance with the current state of art of cQED implementations, as shown in table~\ref{tb:pmt}. Note that the temperature of the bath is chosen such that $\hbar\beta\omega_{\text{q}}=5$, corresponding to $\bar{n}(\omega_{\text{q}})\approx 6.8\times 10^{-3}$. For a typical superconducting-qubit frequency of $\omega_{\text{q}}/(2\pi)=5.0$ GHz, such temperature corresponds to approximately $48$ mK, lying within the achievable temperatures using dilution refrigerators in typical circuit QED setups. Moreover, the Rabi frequency is kept fixed at $1\%$ of the qubit frequency, $\Omega_{\text{d}}/(2\pi)=50$ MHz, and the environment cutoff frequency at $\omega_\textrm{c}/(2\pi)=250$ GHz, which is extended to agree with the high-cutoff approximation $\omega_\textrm{c}\rightarrow\infty$ in the SLED formalism. 

We simulate the qubit dynamics for a broad range of effective qubit dissipation rates $\gamma$, comprising values of $2\pi\times2.5$ MHz ($0.05\%$ of $\omega_{\text{q}}$) up to $2\pi\times500$ MHz ($10\%$ of $\omega_{\text{q}}$). Such tunability has been demonstrated in similar physical setups in recent protocols for engineered environments~\cite{Viitanen2021,Partanen2019,Sevriuk2019,Silveri2019,Partanen2018,Tan2017}. For simplicity of comparison between Lindblad and SLED in the present model, we assume that the intrinsic dephasing and decay rates of the qubit are low compared to $\Omega_{\text{d}}$ and $\gamma$, such that they have a negligible effect on the qubit dynamics. 

We highlight that the parameters associated to the superconducting resonator in table~\ref{tb:pmt} are chosen as reference and do not enter in the numerical simulations of this work. They are only considered in the phenomenological description of the measurement setup implemented for both qubit readout and pump-probe spectroscopy as it will be detailed in section~\ref{sec:mollow}. The large detuning between the bare qubit and resonator frequencies has been extensively used in the dispersive readout of superconducting qubits~\cite{Krantz2019}.

In figure~\ref{fig:dyna}, we find a very good agreement between the SLED and Lindblad methods if the bath-induced energy shift is not neglected and if we use a relatively small value of $\gamma=5\omega_\textrm{q}\times 10^{-3}$. Thus the Born--Markov approximations in this weak-coupling regime seem valid. Such an agreement is manifested by the considerably high value of the corresponding temporally averaged fidelity $\bar{\mathcal{F}}$ shown in figure~\ref{fig:dynb}. 
However, neglecting the bath-induced energy shift alters the dynamics significantly, reducing $\bar{\mathcal{F}}$ by approximately $10\%$. The reduction is even more pronounced in the case $\gamma=\omega_\textrm{q}\times10^{-2}$, where the dissipation rate and the Rabi frequency are equal in magnitude.
Therefore, such a deviation is maximum at the critical ratio $\epsilon_\text{c} \equiv \gamma/\Omega_{\textrm{d}} \approx 1$,
which can be analytically obtained from the asymptotic fidelities between the Lindblad solutions with and without the bath-induced energy shift, see~\ref{subsec:LMEsteady}. Taking such an energy shift into account is naturally addressed beyond the LME~\cite{Thingna2012}. However, we observe quantitatively that it can be relevant for the dynamics of a dissipative driven system even within the weak-coupling approximations. For on-demand dissipation, such a shift needs to be carefully considered in the experiments. 

In addition to giving rise to faster stabilization time scales, the progressive increase of dissipation over the Rabi frequency ($\gamma/\Omega_{\text{d}}>1$) attenuates the relevance of the bath-induced energy shift in the Lindblad dynamics, as shown in figure~\ref{fig:dyna} for $\gamma=\omega_\textrm{q}\times10^{-1}$, and in figure~\ref{fig:dynb} for several different dissipation rates. 
We attribute this intriguing behavior of the dissipative dynamics of the driven qubit to the amount of coherent superposition between the eigenstates of the bare qubit promoted by the drive, which is increased for a resonant drive but inhibited in the strongly dissipative regime. Thus, for the chosen drive frequency $\omega_\textrm{d}=\omega_\textrm{q}$ in figure~\ref{fig:dyn}, the inclusion of the bath-induced energy shift in the LME renders the drive nonresonant with the actual qubit frequency modified by the bath. This in turn leads to a non-vanishing $x$ component of the Bloch vector in the rotating frame. With increasingly strong environmental coupling however, the amount of coherence promoted by the drive field decreases, even in the resonant case, and the decay towards the thermal steady state is favored. This state commutes with $\hat{\sigma}_z$, and hence is dynamically unaffected by the bath-induced energy shift. 

Despite the negligible effect of the bath-induced energy shift on the steady state of the LME in the dissipation-dominated regime $\gamma \gg \Omega_\textrm{d}$, the overall validity of the LME is compromised. Namely, figure~\ref{fig:dynb} shows a progressive reduction of the fidelity $\bar{\mathcal{F}}$ as a function of increasing $\gamma \gtrsim \Omega_\textrm{d}$.
Manifestations of this breakdown are shown for the $z$ component of the Bloch vector in Figure~\ref{fig:dync}, where we observe non-exponential short-time dynamics and a shift in the steady-state values for the SLED method, both being not contemplated by the Born--Markov approximations~\cite{Tuorila2019}. Note that owing to the low bath temperature, the effect of the bath-induced energy shift on the thermal populations in SLED solution is expected to be negligible here.  

\subsection{Steady state: bath-induced energy shift and failure of the Lindblad master equation} \label{subsec:steady}

Let us further detail the importance of the bath-induced energy shift on the Lindblad description of the open dynamics by inspecting the steady-state properties of the driven qubit using equation~\eqref{eq:drho4}.
This serves as a guide for a measure of the inadequacy of the LME for asymptotic predictions, as we introduce below. The use of the SLED is aimed here at simulating the qubit dynamics in a typical cQED experiment.

For convenience, we denote the components of the steady state of the system in the rotating-frame for an arbitrary detuning $\Delta_{\text{q}}=\omega_{\text{q}}+\Delta_{\text{s}}-\omega_{\text{d}}$ as 
\begin{align}
	\sigma_i^{\text{ssf}}(\Delta_{\text{q}})=\Tr[\hsg_{i}\text{e}^{-\text{i}\hsgz\omega_\textrm{d}t/2}\hrh_{\text{S}}^{\text{ss}}\text{e}^{\text{i}\hsgz\omega_\textrm{d}t/2}], \label{eq:sigmaz}
\end{align}
where $i=x,y,z$, $\Delta_{\text{s}}$ is the bath-induced energy shift, and $\hrh_{\text{S}}^{\text{ss}}=\hrh_{\text{S}}(t\rightarrow\infty)$ is the asymptotic density operator of the system in Schr\"odinger's picture. In addition, in the case where the qubit is driven by a weak field ($\Omega_{\text{d}}\ll\omega_{\text{q}}$), the drive Hamiltonian can be written in the rotating-wave approximation such that $\hH_{\text{d}}(t)\approx\hbar\Omega_{\text{d}}(\kb{0}{1}\text{e}^{\text{i}\omega_\text{d}t}+\kb{1}{0}\text{e}^{-\text{i}\omega_\text{d}t})/2$. Consequently, the asymptotic components $\sigma_i^{\text{ssf}}(\Delta_{\text{q}})$ can be found analytically from the LME~\eqref{eq:drho4} for an arbitrary detuning $\Delta_{\text{q}}$. Based on this result, we express how a nonresonant drive modifies the steady state of the qubit in comparison to the resonant-drive case $\Delta_{\text{q}}=0$ by defining the difference
\begin{align}
	\Delta \sigma_i^{\rm ss}=\sigma_i^{\text{ssf,L}}(\Delta_{\text{q}})-\sigma_i^{\text{ssf,L}}(0),\label{eq:DeltaBloch}
\end{align}
where the superscript `L' highlights that such quantities are obtained through the Lindblad equation. As shown in~\ref{subsec:LMEsteady}, we find  
\begin{align}
	\Delta \sigma_x^{\rm ss} & =-\frac{4\left(\frac{\gamma}{\gamma_{\text{\ensuremath{\beta}}}}\right)\left(\frac{\Delta_{\text{q}}}{\Omega_{\text{d}}}\right)}{\left[2+\left(\frac{\gamma_{\beta}}{\Omega_{\text{d}}}\right)^{2}+4\left(\frac{\Delta_{\text{q}}}{\Omega_{\text{d}}}\right)^{2}\right]},\nonumber \\
	\Delta \sigma_y^{\rm ss} & =\frac{8\left(\frac{\gamma}{\Omega_{\text{d}}}\right)\left(\frac{\Delta_{\text{q}}}{\Omega_{\text{d}}}\right)^{2}}{\left[2+\left(\frac{\gamma_{\beta}}{\Omega_{\text{d}}}\right)^{2}+4\left(\frac{\Delta_{\text{q}}}{\Omega_{\text{d}}}\right)^{2}\right]\left[2+\left(\frac{\gamma_{\beta}}{\Omega_{\text{d}}}\right)^{2}\right]},\nonumber \\
	\Delta \sigma_z^{\rm ss} & =\frac{8\left(\frac{\gamma}{\gamma_{\beta}}\right)\left(\frac{\Delta_{\text{q}}}{\Omega_{\text{d}}}\right)^{2}}{\left[2+\left(\frac{\gamma_{\beta}}{\Omega_{\text{d}}}\right)^{2}+4\left(\frac{\Delta_{\text{q}}}{\Omega_{\text{d}}}\right)^{2}\right]\left[2+\left(\frac{\gamma_{\beta}}{\Omega_{\text{d}}}\right)^{2}\right]}.\label{eq:DeltaBloch2}
\end{align}
where $\gamma_{\beta}=\gamma[2\bar{n}(\omega_\text{q})+1]$. Therefore, the quantities $\Delta \sigma_i^{\rm ss}$ indicate the LME predictions for the qubit sensitivity on the frequency change of a weak and monochromatic transverse drive.
\begin{figure*}[t]
	\subfloat{\label{fig:steadya}} 
	\subfloat{\label{fig:steadyb}}
	\subfloat{\label{fig:steadyc}} 
	\centering
	\includegraphics[width=\linewidth]{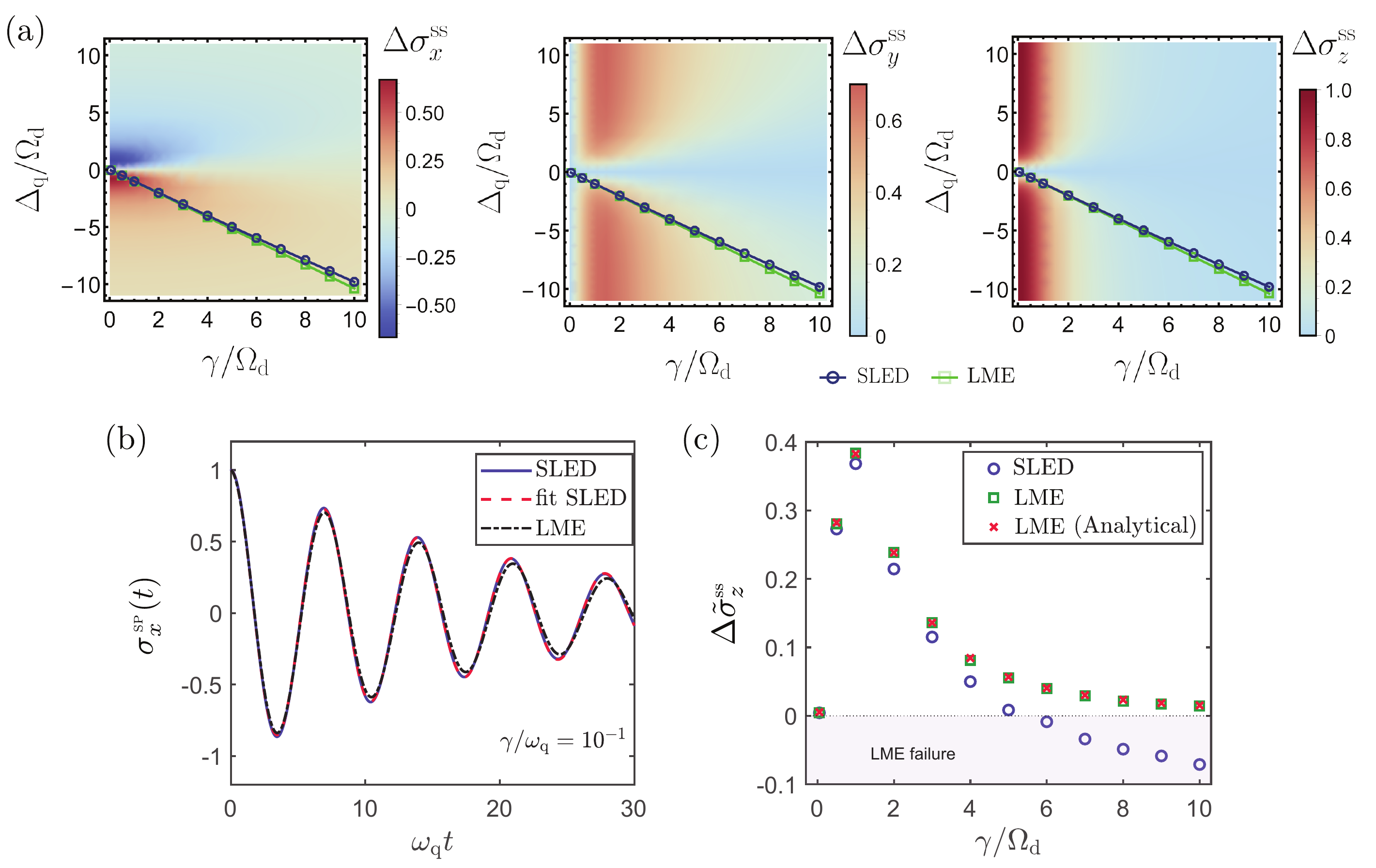}
	\caption{(a) Analytically obtained difference of the steady-state component of the system Bloch vector from that in the case of resonant drive, $\Delta \sigma_i^{\rm ss}$ ($i=x,y,z$) given by equations~\eqref{eq:DeltaBloch2}, as a function of the dissipation rate $\gamma$ and the detuning of the drive angular frequency from resonance $\Delta_{\text{q}}$. Assuming that the drive is set to the angular frequency of the bare qubit, the bath-induced energy shift induces a finite detuning $\Delta_{\text{q}}$ as shown for SLED (blue markers) and for the LME (green markers).
		(b) The $x$ component of the Bloch vector in the Sch\"odinger picture as a function of time for SLED (solid line) and LME (dash-dotted line) simulations without drive ($\Omega_\textrm{d}=0$). The qubit is initially prepared in the eigenstate of $\hsgx$ with eigenvalue~1. We fit an exponentially damped cosine function (dashed line) to the SLED solution. The obtained oscillation frequency corresponds to the qubit frequency modified by the bath which is used in (a) for the SLED data.  
		(c) Difference $\Delta \tilde{\sigma}_z^{\rm ss}$, defined in equation~\eqref{eq:Deltaz}, with the choice $\Delta_\textrm{q}=\Delta_\textrm{s}$ for SLED (circles), numerically solved Lindblad (squares), and Lindblad with rotating-wave approximation (crosses) as a function of the qubit dissipation rate, $\gamma$. The colored region corresponding to $\Delta \tilde{\sigma}_z^{\rm ss}<0$ indicates the failure of LME as expected from the definition in equation~\eqref{eq:Deltaz}. All qubit parameters not given here are fixed according to table~\ref{tb:pmt}.}\label{fig:steady}
\end{figure*}

In figure~\ref{fig:steadya}, we show the dependence of $\Delta \sigma_i^{\rm ss}$ on the qubit decay rate $\gamma$ and on the angular frequency detuning of the drive $\Delta_{\text{q}}$ for a low-temperature environment ($\gamma_{\beta}\approx \gamma$). A non-resonant drive on the qubit ($\Delta_{\text{q}}\neq0$) affects the different components of the Bloch vector in different ways. The difference in the $x$ component changes its sign with that of the detuning and is clearly pronounced in the region of moderate dissipation, tending to vanish at $|\Delta_{\text{q}}/\Omega_{\text{d}}|\gg 1$. On the other hand, the detuning barely affects the $y$ component for $\gamma/\Omega_{\text{d}}\ll1$, and in this regime, $\Delta \sigma_z^{\rm ss}$ saturates to a high value for a sufficiently large $|\Delta_{\text{q}}/\Omega_{\text{d}}|$. In general, the difference in all components of the Bloch vector decreases and becomes independent of the detuning with increasing $\gamma/\Omega_{\text{d}}\gg 1$, which is another manifestation of dissipation dominating over the drive dynamics.

Provided that the drive frequency is set to $\omega_{\text{d}}=\omega_{\text{q}}$,
we have $\Delta_{\text{q}}=\Delta_{\text{s}}$. For this choice of $\Delta_{\text{q}}$, the markers in figure~\ref{fig:steadya} show the bath induced energy-shift as function of the qubit decay rate. Whereas the perturbative approach of the dissipative dynamics allows one to obtain $\Delta_{\text{s}}$ directly from equation~\eqref{eq:LS1}, we obtain it for the SLED method by fitting an exponentially damped cosine function to the early decay of the qubit coherence as illustrated in figure~\ref{fig:steadyb}. For the chosen parameters, the relation between $\Delta_{\text{s}}$ and $\gamma$ is well approximated by a linear fit in both methods. As shown in \cite{Tuorila2019}, this dependence ceases to be linear for strong enough system--bath coupling strength. 

Interestingly, equations~\eqref{eq:DeltaBloch2} along with figure~\ref{fig:steadya} show that $\Delta \sigma_y^{\rm ss}$ and $\Delta \sigma_z^{\rm ss}$ are symmetric with respect to $\Delta_{\text{q}}$.
Focusing our attention to the $z$ component, we observe that $\Delta \sigma_z^{\rm ss}\geq 0$ in equations~\eqref{eq:DeltaBloch2} for any choice of parameters. 
Based on this result, we introduce the measure
\begin{align}
	\Delta \tilde{\sigma}_z^{\rm ss}={\sigma}_{z}^{\text{ssf}}(\Delta_{\text{q}})-\sigma_{z}^{\text{ssf,L}}(0), \label{eq:Deltaz}
\end{align}
where $\sigma_{z}^{\text{ssf,L}}(0)$ is the steady-state $z$ component of the Bloch vector given by the LME at resonance and ${\sigma}_{z}^{\text{ssf}}(\Delta_{\text{q}})$ is obtained by our method of choice or even experimentally. Note that if  ${\sigma}_{z}^{\text{ssf}}(\Delta_{\text{q}})$ is also obtained from the LME, equation~\eqref{eq:Deltaz} reduces to $\Delta \sigma_z^{\rm ss}$ defined in equations~\eqref{eq:DeltaBloch2}, which is always positive. Thus, this measure can be used to identify regimes where the perturbative approach encoded in the LME is not sufficient to correctly predict the steady state of the weakly driven qubit.  
In an experiment, $\sigma_{z}^{\text{ssf,L}}(0)$ can be inferred from the characterization of parameters involved in the dynamics combined with a subsequent analytical calculation of the asymptotic $z$ component [see equation~\eqref{eq:BlochVss} of~\ref{subsec:LMEsteady}]. On the other hand, ${\sigma}_{z}^{\text{ssf}}(\Delta_{\text{q}})$ can be obtained through usual steady-state readout of the qubit driven out of resonance.
The negativity of $\Delta \tilde{\sigma}_z^{\rm ss}$ violates the lower bound imposed by equation~\eqref{eq:DeltaBloch2}, being a sufficient condition for the failure of the time-local LME.

Figure~\ref{fig:steadyc} shows $\Delta \tilde{\sigma}_z^{\rm ss}$ for selected values of $\gamma/\Omega_{\text{d}}$ at $\Delta_{\text{q}}=\Delta_{\text{s}}$. The values of $\sigma_{z}^{\text{ssf}}(\Delta_{\text{q}})$ are calculated from long-time solutions of the LME and SLED, which are intended to simulate the dynamics of the qubit in an experiment. The good agreement between equation~\eqref{eq:DeltaBloch2} and the numerical results from the Lindblad equation highlights the validity of the RWA in the drive Hamiltonian $\hH_{\text{d}}(t)$. In these cases, as expected from equation~\eqref{eq:DeltaBloch2}, $\Delta \tilde{\sigma}_z^{\rm ss}$ is positive for all decay rates and achieves its maximum for $\gamma\approx\Omega_{\text{d}}$. However, the detuned asymptotic $z$ component given by SLED produces $\Delta \tilde{\sigma}_z^{\rm ss}<0$ for dissipation rates $\gamma>0.05\times\omega_{\text{q}}$, or in terms of the parameters of table~\ref{tb:pmt}, for $\gamma>2\pi\times250$ MHz. This is a clear evidence of incompatibility with the used weak-coupling approximations.

As pointed out in Ref.~\cite{Tuorila2019} for a non-driven qubit, the shift of $\sigma_z(t)$ 
given by the SLED compared to that by the Lindblad equation cannot be fully attributed to a bath-induced energy shift since the correlations between the qubit and the bath created during the dynamics contributes as well. By turning on a very weak drive field, one is potentially able to study threshold conditions where $\Delta \tilde{\sigma}_z^{\rm ss}=0$, indicating that such correlations may be significant. From a different perspective, our measure serves as a fine benchmark of the time-local Lindblad equation for a weakly driven qubit, thus shedding light on the validity limits of weak-coupling assumptions on the open dynamics. 
However, the threshold is still relaxed in the sense that possible deviations owing to the weak-coupling assumptions are not directly detected if $\Delta\tilde{\sigma}_z^{\rm ss}>0$ even though they tend to increase with $\gamma$ in our particular case as shown in figure~\ref{fig:steadyc}. This observation clearly illustrates the practical relevance of using non-perturbative approaches for accurate predictions.
	\section{Pump-probe spectroscopy} \label{sec:mollow}

The second example of a driven dissipative system presented in this work consists of a qubit driven by a bichromatic field of the form
\begin{align}
	f(t)=\Omega_{\text{d}}\cos(\omega_{\text{d}}t)+\Omega_{\text{p}}\cos(\omega_{\text{p}}t+\pi/2),\label{eq:f2}
\end{align}
which is a sum of the monochromatic drive field of equation~\eqref{eq:f1}, referred to as the primary drive, and a probe field with angular frequency $\omega_{\text{p}}$ and an associated Rabi angular frequency $\Omega_{\text{p}}$.
Below, we employ the SLED and compare it with the Lindblad formalism to study the signatures of the qubit fluorescence spectrum by means of a pump-probe approach \cite{Baur2009}. Specifically, assuming $\Omega_{\text{p}}\ll\Omega_{\text{d}}$  so that the probe field acts as a weak perturbation to the driven qubit, information about the spectrum under the primary drive is obtained from the response of the system to the probe as the angular frequency $\omega_{\text{p}}$ is swept.

Rather than monitoring the radiation spectrum of the qubit, we study the response of the system through the temporally averaged $z$ component of the Bloch vector  
\begin{align}
	\bar{\sigma}_z=\frac{1}{n_{\text{p}}t_{\text{p}}}\int_{t_\text{f}-n_{\text{p}}t_{\text{p}}}^{t_\text{f}}\sigma_z(t)\textrm{d}t, \label{eq:tavgz}
\end{align}  
where $\sigma_z(t)=\Tr[\hsg_z\hrh_{\text{S}}(t)]$ as above, $n_{\text{p}}t_{\text{p}}$ is the length of the integration interval, and $t_\text{f}$ is the final time chosen such that the initial transient dynamics has a negligible effect on $\bar{\sigma}_z$. 
If the probe and the drive are out of resonance, $\sigma_z(t)$ tends to oscillate in time with an amplitude $h_z$ and frequency $|\Delta_{\text{p}}|\approx|\omega_{\text{p}}-\omega_{\text{d}}|$, so that the average in equation~\eqref{eq:tavgz} is calculated over multiple integers $n_{\text{p}}$ of its period $t_{\text{p}}=2\pi/|\Delta_{\rm p}|$. At resonance $\Delta_{\rm p}=0$, the temporal dependence of $\sigma_z(t)$ is negligible due to the single oscillation frequency in equation~\eqref{eq:f2} and the considerably small chosen values of $\Omega_{\rm d}$ and $\Omega_{\rm p}$ compared to the qubit angular frequency (see table~\ref{tb:pmt}).

In a typical cQED experiment, one can relate $\bar{\sigma}_z$ with the field transmitted through a readout resonator dispersively coupled to the qubit \cite{Krantz2019}. In the semiclassical approximation \cite{Tuorila2009}, a phenomenological inclusion of the resonator yields for the asymptotic field amplitude transmitted from the readout resonator to its output port (see~\ref{sec:app2}) 
\begin{align}
	A = \frac{\Omega_{\text{m}}}{\kappa}\frac{1}{\sqrt{1+\left(\frac{2\chi \bar{\sigma}_z}{\kappa}\right)^2}}, \label{eq:Amp}    
\end{align}
where $\Omega_{\text{m}}$ is the amplitude of a weak measurement drive continuously applied on the input port of the resonator, $\kappa$ is the resonator energy decay rate that is assumed to be dominated by leakage to the output port, 
and $\chi$ is the so-called dispersive shift associated to the qubit-resonator coupling.
\begin{figure*}[t]
	\subfloat{\label{fig:mollowa}} 
	\subfloat{\label{fig:mollowb}}
	\centering
	\includegraphics[width=\linewidth]{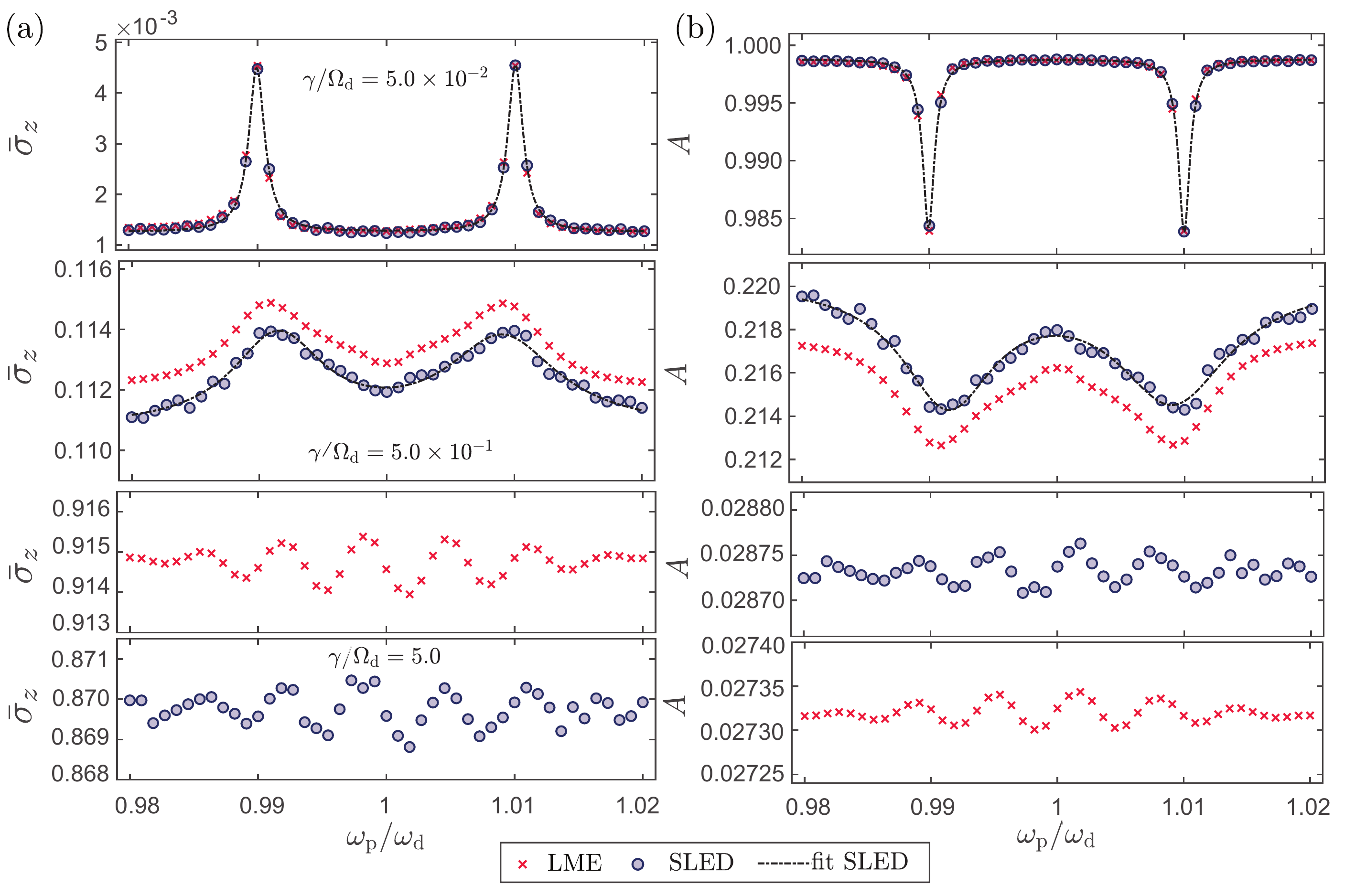}
	\caption{(a) Temporally averaged $z$ component of the steady-state Bloch vector $\bar{\sigma}_z$ as a function of the probe frequency $\omega_{\text{p}}$ for different values of the qubit decay rate $\gamma$. (b) Corresponding field amplitude $A$ transmitted from the qubit readout resonator to its output port. The drive angular frequency $\omega_\textrm{d}$ is chosen to match the qubit frequency including any bath-induced energy shifts. 
		In the top and center panels, two Lorentzians are fitted (dash-dotted line) to the SLED data (dots), thus indicating the sidebands of the Mollow triplet. The parameters not given here are chosen as in table~\ref{tb:pmt}.}\label{fig:mollow}
\end{figure*}

Figure~\ref{fig:mollow} shows $\bar{\sigma}_z$ and $A$ as functions of the probe frequency for various dissipation rates $\gamma$. Similar to section~\ref{sec:monofield}, the parameters are chosen according to table~\ref{tb:pmt} unless otherwise stated. Here, the drive frequency $\omega_{\text{d}}$ is adjusted to the resonance with the frequency of the qubit including any bath-induced energy shifts for each $\gamma$. The bath-induced frequency shift is calculated as in section~\ref{subsec:steady}, that is, through equation~\eqref{eq:LS1} for the Lindblad master equation and through a fit to the damped decay of the qubit coherence for the SLED. 

We show in figure~\ref{fig:mollow} that for very weak dissipation there is a good agreement between the two methods for $\bar{\sigma}_z$ and $A$. As $\gamma$ increases, the 
solutions given by the two approaches tend to separate,
indicating that the steady state of the Lindblad master equation significantly deviates from the one given by SLED. Despite numerical fluctuations caused by the finite number of noise trajectories used for SLED, we observe that some resonance-like features tend to be preserved even for the dissipation rate of the order of the Rabi angular frequency of the drive. The main differences arise from the probe-frequency-independent shift of the response. Similar to section~\ref{sec:monofield}, this effect is caused by a shift of $\sigma_z(t)$ given by SLED as compared to that produced by the Lindblad master equation, becoming more pronounced as $\gamma$ increases. In contrast to figure~\ref{fig:dyn}, in which the drive frequency is fixed at the bare qubit frequency in both methods ($\omega_{\text{d}}=\omega_{\text{q}}$), the shift appears in figure~\ref{fig:mollow} for drive frequencies $\omega_{\text{d}}$ matching the qubit transition frequency shifted by the bath ($\omega_{\text{d}}=\omega_{\text{q}}+\Delta_{\text{s}}$). This suggests that such a phenomenon is not primarily caused by the renormalization of the qubit frequency, or any unitary effect, but rather it may be attributed to the appearance of asymptotic system--bath correlations as the strong coupling is approached.

A qualitative analysis may connect the presented results with the actual qubit fluorescence spectrum predicted by Mollow in Ref.~\cite{Mollow1969}. Typically, the radiation spectrum is proportional to the Fourier transform of a two-time correlation function of the system evaluated at its steady state, e.g., $R(\omega)=\int_{-\infty}^{\infty}\text{d}\tau\, \text{e}^{-\text{i}\omega\tau}\ev{\hdgg{C}(\tau)\hC(0)}_{\text{ss}}$, with $\hC=\kb{0}{1}$ being an example in the case of a qubit. For a weak environmental coupling, the calculation of such correlation functions is usually obtained through the quantum regression theorem \cite{Carmichael2013,Heinz-PeterBreuer2007}, where one resorts to the Born--Markov approximations. Within this approach, for  $\gamma/\Omega_{\text{d}}\ll 1$, the fluorescence spectrum of a dissipative qubit that is driven by a resonant field of the form of equation~\eqref{eq:f1} in the RWA presents three peaks centered at frequencies  
$\omega_0=\omega_{\text{d}}$ and $\omega_{\pm}=\omega_{\text{d}}\pm \Omega_{\text{d}}$ ~\cite{Mollow1969}. 
The sideband peaks have a Lorentzian shape that becomes broadened and flattened as the ratio $\gamma/\Omega_{\text{d}}$ increases. These features are present in the top and middle panels of figure~\ref{fig:mollowa}, where we fit the data provided by the SLED with Lorenztian functions peaked roughly at $\omega_{-}=0.99\times\omega_{\text{d}}$ and $\omega_{+}=1.01\times\omega_{\text{d}}$. In the bottom panels, however, the sideband peaks are absent due to the high qubit dissipation rate. The small oscillations in this case, which are noticeable in both the SLED and Lindblad data, may be attributed to the different number of periods $n_{\text{p}}$ used in the numerical integration of equation~\eqref{eq:tavgz}.
\begin{figure}[t]
	\centering
	\includegraphics[width=0.4\linewidth]{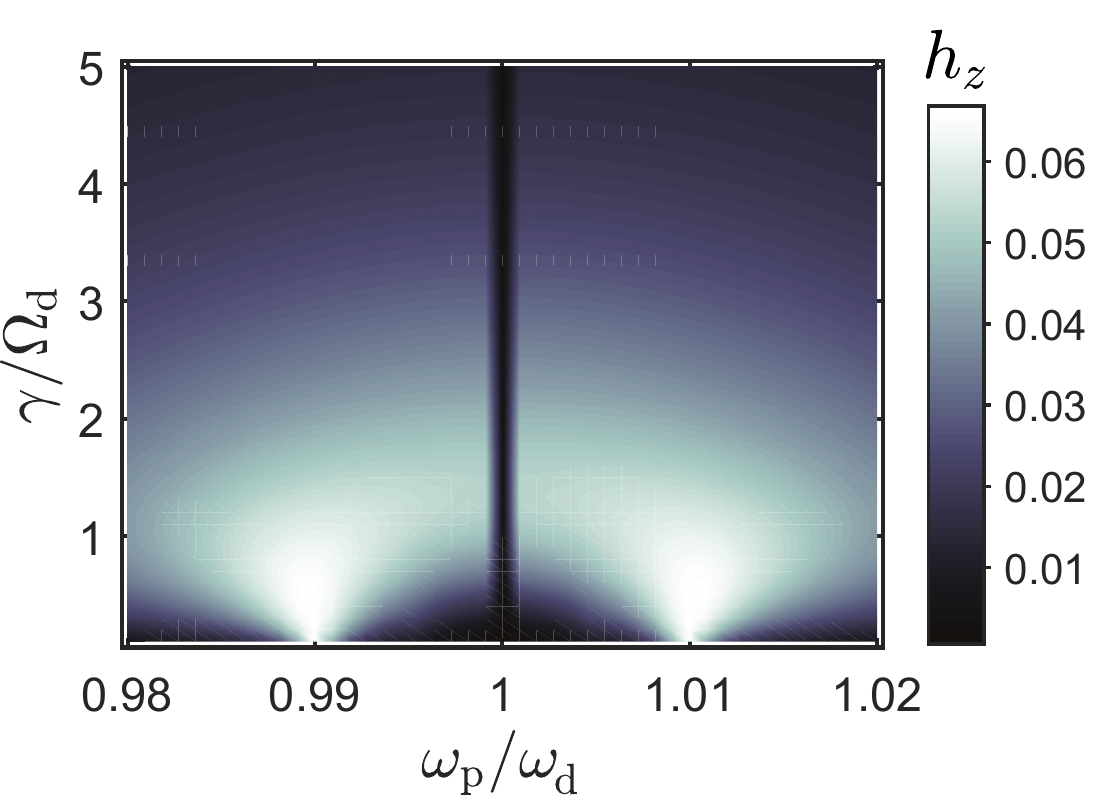}
	\caption{Amplitude $h_z$ of the probe-induced oscillation of $\sigma_z(t)$ as a function of the probe frequency $\omega_{\text{p}}$ and the qubit dissipation rate $\gamma$. Data is obtained through the Lindblad master equation~\eqref{eq:drho4} with $f(t)$ from equation~\eqref{eq:f2}. The parameters not given here are chosen as in table~\ref{tb:pmt}.}\label{fig:hz}
\end{figure}

In contrast to the studies of Ref.~\cite{Mollow1969}, no central peak at $\omega_0=\omega_{\text{d}}$ is observed in figure~\ref{fig:mollow}, since the probe field does not excite the qubit at resonance, according to the definition of $f(t)$ in equation~\eqref{eq:f2}. This can also be checked by writing the driven qubit Hamiltonian in the frame rotating at the primary drive frequency. Since the pump-probe approach renders the long-time behavior of $\sigma_z(t)$ an indicator of the presence of the probe field, the latter is not perceived by the qubit when $\Delta_{\text{p}}=0$. Alternatively, this and the above-mentioned qualitative features of the qubit fluorescence spectrum can be checked in figure~\ref{fig:hz}, where we show the amplitude $h_z$ of the probe-induced oscillations of $\sigma_z(t)$ as a function of the probe frequency $\omega_\text{p}$ and a broad range of qubit dissipation rates $\gamma$. We clearly observe that the regions of high amplitude indicate the sideband peaks of the Mollow triplet, these being pronounced and narrow at small $\gamma/\Omega_{\text{d}}$. These peaks become flat and broad at high qubit dissipation rates, eventually disappearing at $\gamma/\Omega_{\text{d}}\gg 1$. 

Similar damped oscillations as shown here have also been observed in different physical setups, for instance, in two coupled degenerate resonators with significantly different leakage rates~\cite{Partanen2019}. In addition, the radiation spectrum of a qubit under a bichromatic field in the RWA has also been obtained through the quantum regression theorem and presents a rich variety of phenomena depending on the choice of parameters \cite{Agarwal1991}. However, the features of the qubit spectrum under the drive field of equation~\eqref{eq:f1} are preserved assuming that it is much stronger than the probe field, the case considered in this work.

	\section{Conclusions} \label{sec:conc}
We assessed the precision of weak-coupling assumptions of a driven qubit interacting with a bosonic environment through examples where the non-perturbative stochastic Liouville--von Neumann equation, or SLED, is appropriate. We focused our attention on the case where the qubit interacts with weak and nearly resonant transverse fields along with a cold Ohmic bath with low dissipation rates compared with the bare system frequency. Such a scenario is typically addressed by the Lindblad master equation, or LME, and it is of practical relevance in state-of-art implementations of engineered environments in circuit quantum electrodynamics aimed, for example, at optimized initialization protocols of the system. Thus, our investigation complements the recent studies published in~\cite{Tuorila2019} and~\cite{Vadimov2021} on the benchmark of SLED over perturbative master equations.

We carried out a quantitative comparison of SLED with the LME and showed that the often overlooked bath-induced energy shift in the LME becomes less relevant for the dynamics with the strength of the dissipation increasing well beyond the drive Rabi frequency. However, new effects arising from the failure of the weak-coupling assumptions emerge in these regimes, being captured by the non-perturbative treatment of the drive and dissipation given by the SLED. In addition, we proposed a measure based on the sensitivity of the qubit population to the drive frequency. As a consequence, we identified regimes where the SLED yields for the steady state of the qubit dynamics distinctive and quantitatively measurable differences to the results of the Lindblad equation. Moreover, we have used the SLED and Lindblad approaches to study the signatures of the qubit fluorescence spectrum for different dissipation rates that may be produced by tunable environments in cQED. 

In conclusion, our results may guide future experiments to probe driven open quantum systems and the validity of the weak-coupling approximations in describing their dynamics. This potentially allows for the exploration of undiscovered frontiers which are not well captured by the weak-coupling Markovian dynamics. In particular, our work may motivate further investigations on the validity of other perturbative master equations, such as the Floquet--Born--Markov equation, where the driving has been more accurately taken into account in the derivation of the master equation. However, despite improvements arising from deriving the dissipators in the dressed state basis, such equations are nevertheless perturbative. As a consequence, a precise proposal for experiments and corresponding parameters in scenarios of increasing dissipation as presented in this work calls for a model contemplating both the drive--dissipation interplay and high-order corrections to the system--bath correlations, as given by the SLED. A more transparent comparison of the SLED with other non-perturbative approaches in the context of circuit quantum electrodynamics emerges as natural future line of research.

	\section*{Acknowledgements} \label{sec:acknowledgements}
We acknowledge Funda\c c\~ao de Amparo \`a Pesquisa
do Estado de S\~ao Paulo (FAPESP) through grants No.~2017/09058-2 and No.~2018/26726-1, the Brazilian National Institute of Science and Technology of Quantum Information (CNPq INCT-IQ 465469/2014-0), CNPq (grant No.~305723/2020-0), CAPES/PrInt (88881.310346/2018-01), the Academy of Finland under its Centres of Excellence Program grants No.~312300 and No.~336810, the European Research Council under grants Nos.~681311 (QUESS) and 957440 (SCAR), the Jane and Aatos Erkko Foundation, and the Technology Industries of Finland Centennial Foundation for financial support.
The authors also acknowledge CSC - IT Center for Science, Finland, for generous computational resources, and Joachim Ankerhold, J\"urgen Stockburger, Tapio Ala-Nissil\"a, Aravind Babu, and Sahar Alipour for discussions.
	
	\appendix
	\section{Lindblad master equation} \label{sec:Lindblad}

Below, we review the derivation of the Lindblad master equation for the driven dissipative qubit described in section~\ref{sec:model} of the main text. Similar derivations have been reported in the existing literature~\cite{Mollow1969a,Carmichael2013,Heinz-PeterBreuer2007}, and hence the discussion in this Appendix is given mainly for the sake of completeness of our notation.

We begin by employing the interaction picture with respect to the free Hamiltonian $\hH_{\text{S}}+\hH_{\text{B}}$, such that the exact temporal evolution of the total density operator $\hrh^\text{i}(t)$ is given by the Liouville--von Neumann equation $\textrm{d}\hrh^\text{i}(t)/\textrm{d}t=-\text{i}[\hH^\text{i}(t),\hrh^\text{i}(t)]/\hbar$, where $\hH^\text{i}(t)=\hH^\text{i}_{\text{d}}(t)+\hH^\text{i}_{\text{SB}}(t)$ and the superscript~i stands for the interaction picture. A recursive integration up to the second order and a trace over the bath degrees of freedom yield
\begin{align}
\dert{\hrh_{\text{S}}^\text{i}(t)}= & -\frac{\text{i}}{\hbar}\left[\hH_{\text{d}}^\text{i}(t),\hrh_{\text{S}}^\text{i}(0)\right]-\frac{1}{\hbar^2}\int_{0}^{t}\textrm{d}t'\left[\hH_{\text{d}}^\text{i}(t),\left[\hH_{\text{d}}^\text{i}(t'),\hrh_{\text{S}}^\text{i}(t')\right]\right]\nonumber \\
&-\frac{1}{\hbar^2}\int_{0}^{t}\textrm{d}t'\,\Tr_{\text{B}}\left\{ \left[\hH_{\text{SB}}^\text{i}(t),\left[\hH_{\text{SB}}^\text{i}(t'),\hrh^\text{i}(t')\right]\right]\right\}\nonumber \\
 & -\frac{1}{\hbar^2}\int_{0}^{t}\textrm{d}t'\,\Tr_{\text{B}}\left\{ \left[\hH_{\text{d}}^\text{i}(t),\left[\hH_{\text{SB}}^\text{i}(t'),\hrh^\text{i}(t')\right]\right]\right\} \nonumber \\
 & -\frac{1}{\hbar^2}\int_{0}^{t}\textrm{d}t'\,\Tr_{\text{B}}\left\{ \left[\hH_{\text{SB}}^\text{i}(t),\left[\hH_{\text{d}}^\text{i}(t'),\hrh^\text{i}(t')\right]\right]\right\},\label{eq:drho1}
\end{align}
where $\hrh_{\text{S}}^\text{i}(t)=\Tr_{\text{B}}[\hrh^\text{i}(t)]$
is the reduced density operator of the system at time instant $t$ and we have assumed an initially factorized state as mentioned in the main text. Since the first moments of observables calculated in the state $\hrh_{\text{B}}^\text{i}(0)$ may be chosen to vanish, one naturally obtains $\Tr_{\text{B}}[\hH_{\text{SB}}^\text{i}(t),\hrh^\text{i}(0)]=0$, which is hence not visible in equation~\eqref{eq:drho1}. In the absence of the drive [$f(t)=0$], only the third term on the right side of equation~\eqref{eq:drho1} remains. 

The structure of equation~\eqref{eq:drho1} is rather complicated as it is an integro-differential exact equation. However, it can be simplified under a series of assumptions. First, one writes the total density operator as $\hrh^\text{i}(t)=\hrh_{\text{S}}^\text{i}(t)\otimes\hrh_{\text{B}}^\text{i}(0)+\hom^\text{i}(t)$,
where $\hom^\text{i}(t)$ is present only when system--bath correlations are created during the dynamics. Naturally, $\Tr[\hom^\text{i}(t)]=0$ in order to preserve the normalization. 
The so-called Born approximation physically asserts that the influence of the system on the bath dynamics is small so that it essentially stays in the Gibbs state throughout the interaction. Consequently, one neglects the correlation term $\hom^\text{i}(t')$ when considering $\hrh^\text{i}(t')$ in equation~\eqref{eq:drho1}, and hence we obtain
\begin{align}
\dert{\hrh_{\text{S}}^\text{i}(t)}= & -\frac{\text{i}}{\hbar}\left[\hH_{\text{d}}^\text{i}(t),\hrh_{\text{S}}^\text{i}(0)\right]-\frac{1}{\hbar^2}\int_{0}^{t}\textrm{d}t'\left[\hH_{\text{d}}^\text{i}(t),\left[\hH_{\text{d}}^\text{i}(t'),\hrh_{\text{S}}^\text{i}(t')\right]\right] \nonumber \\
& -\frac{1}{\hbar^2}\int_{0}^{t}\textrm{d}t'\Tr_{\text{B}}\left\{ \left[\hH_{\text{SB}}^\text{i}(t),\left[\hH_{\text{SB}}^\text{i}(t'),\hrh_{\text{S}}^\text{i}(t')\hrh_{\text{B}}^\text{i}(0)\right]\right]\right\} .\label{eq:drho2}
\end{align}
Note that the Born approximation implemented inside the the double commutators of equation~\eqref{eq:drho1} does not guarantee the conservation of the system entropy as in a unitary evolution. 

We note that in equation~\eqref{eq:drho2}, the two last terms in
equation~\eqref{eq:drho1} have been dropped. This is a consequence of the Born approximation, i.e., $\hom^\text{i}(t)\rightarrow 0$. Therefore, the Born approximation does not account for the drive--dissipation interplay promoted by such terms in the case of linear interaction of the system with a heat bath. The overlooking of such an interplay also allows one to combine the first two terms of equation~\eqref{eq:drho2} into the single commutator $-\text{i}[\hH_{\text{d}}^\text{i}(t),\hrh_{\text{S}}^\text{i}(t)]/\hbar$ so that the drive contributes only to the unitary dynamics of the qubit. 

The last term of equation~\eqref{eq:drho2} is here simplified by assuming that the system dynamics is memoryless and by coarse-graining in time (Markov approximation). The joint effect of these approximations allows one to neglect the dependence of the state of the system on its past history such that $\hrh_{\text{S}}^\text{i}(t')\rightarrow\hrh_{\text{S}}^\text{i}(t)$ during the time integration, and to extend the integration limit up to infinity. This is usually justified as long as the bath autocorrelation function~\eqref{eq:L1} arising from the double commutator decays faster than the relaxation time of the system. Upon the change of variable $t'\rightarrow t-\tau$, these approximations lead to the master equation   
\begin{align}
\dert{\hrh_{\text{S}}^\text{i}(t)}= & -\frac{\text{i}}{\hbar}\left[\hH_{\text{d}}^\text{i}(t),\hrh_{\text{S}}^\text{i}(t)\right] \nonumber \\
&-\frac{1}{\hbar^2}\int_{0}^{\infty}\textrm{d}\tau\Tr_{\text{B}}\left\{ \left[\hH_{\text{SB}}^\text{i}(t),\left[\hH_{\text{SB}}^\text{i}(t-\tau),\hrh_{\text{S}}^\text{i}(t)\hrh_{\text{B}}^\text{i}(0)\right]\right]\right\} .\label{eq:drho3}
\end{align}
In general, the double commutator in equation~\eqref{eq:drho3} gives rise to a correction of the system energy and non-unitary dynamics. However, a master equation of the form~\eqref{eq:drho3} typically does not generate a completely positive map. One can overcome this problem, and subsequently write equation~\eqref{eq:drho3} in the so-called Lindblad form, by removing fast oscillating terms. This is referred as the secular approximation and it is usually justified in the weak system--bath coupling regime. For the case under study, such terms oscillate according to $\text{e}^{\pm 2\text{i} \omega_{\text{q}}t}$ and do not contribute to energy shifts. Consequently, by carrying out the integration on the right side of equation~\eqref{eq:drho3}, using the secular approximation, and returning to the Schrödinger picture, one obtains the Lindblad master equation~\eqref{eq:drho4} of the main text.

\subsection{Steady-state and fidelity between Lindblad solutions}\label{subsec:LMEsteady}
Here, we show the analytical expressions for the components of the steady-state Bloch vector in the rotating frame according to the Lindblad equation, $\sigma_{i}^{\text{ssf,L}}(\Delta_{\text{q}})$, which can be found following the procedure described in section~\ref{subsec:steady}. They read
\begin{align}
    \sigma_{x}^{\text{ssf,L}}\left(\Delta_{\text{q}}\right)&=-\frac{4\gamma\Omega_{\text{d}}\Delta_{\text{q}}}{\gamma_{\beta}(2\Omega_{\text{d}}^{2}+\gamma_{\beta}^{2}+4\Delta_{\text{q}}^{2})},\nonumber\\
    \sigma_{y}^{\text{ssf,L}}\left(\Delta_{\text{q}}\right)&=-\frac{2\gamma\Omega_{\text{d}}}{2\Omega_{\text{d}}^{2}+\gamma_{\beta}^{2}+4\Delta_{\text{q}}^{2}},\nonumber\\
    \sigma_{z}^{\text{ssf,L}}\left(\Delta_{\text{q}}\right)&=\frac{\gamma(\gamma_{\beta}^{2}+4\Delta_{\text{q}}^{2})}{\gamma_{\beta}(2\Omega_{\text{d}}^{2}+\gamma_{\beta}^{2}+4\Delta_{\text{q}}^{2})}. \label{eq:BlochVss}
\end{align}
Using equations~\eqref{eq:BlochVss} in the definition for $\Delta \sigma_i^{\rm ss}$ in equation~\eqref{eq:DeltaBloch}, yields equations~\eqref{eq:DeltaBloch2} of the main text.

One may also be interested in the fidelity between the steady states of the Lindblad master equation for the qubit driven at the bare frequency ($\Delta_{\text{q}}=\Delta_{\text{s}}$) and driven at the frequency shifted by the system-bath interactions ($\Delta_{\text{q}}=0$). Such a fidelity can be obtained analytically by writing the steady states in terms of the Bloch vector components in equation~\eqref{eq:BlochVss} and using the definition in equation~\eqref{eq:fid}. Assuming $\gamma_{\beta}\approx \gamma$, we find
\begin{align}
    \mathcal{F}(\Delta_{\text{s}})=1-\frac{4\Delta_{\text{s}}^{2}\Omega_{\text{d}}^{2}}{(2\Omega_{\text{d}}^{2}+\gamma^{2}+4\Delta_{\text{s}}^{2})(2\Omega_{\text{d}}^{2}+\gamma^{2})}. \label{eq:fidL}
\end{align}
As shown in figure~\ref{fig:steadya} and from a direct evaluation of equation~\eqref{eq:LS1}, the energy shift $\Delta_{\text{s}}$ in the Lindblad master equation is negative for the chosen parameters in table~\ref{tb:pmt} and depends linearly on $\gamma$. By writing $\Delta_{\text{s}}=-\alpha \gamma$, with $\alpha>0$, it is possible to show that the fidelity in equation~\eqref{eq:fidL} is minimized at the critical ratio
\begin{align}
    \frac{\gamma}{\Omega_{\text{d}}}=\epsilon_\text{c}=\frac{\sqrt{2}}{(1+4\alpha^2)^{1/4}}. \label{eq:gammac}
\end{align}
For values of $\alpha\approx 1$, one obtains $\epsilon_\text{c}\approx 1$. This condition can be qualitatively observed in figures~\ref{fig:dynb} and~\ref{fig:steadyc} of the main text.

\section{Stochastic Liouville-von Neumann equation for dissipation (SLED)} \label{sec:SLED}

Here, we briefly review the main features of the so-called SLED formalism, in the derivation of which one utilizes the path integral description of quantum mechanics \cite{Feynman1963}. Consider a closed quantum system with total position and momentum operators $\hat{\mathbf{q}}$ and $\hat{\mathbf{p}}$, respectively, and the corresponding eigenstates $\ket{\mathbf{q}}$ and $\ket{\mathbf{p}}$. Under the influence of a temporally dependent Hamiltonian $\hH(t)$, the propagator associated to the closed dynamics $K(\mathbf{q}_\text{f},\mathbf{q}_\text{i})=\mel{\mathbf{q}_\text{f}}{\mathcal{T}\text{e}^{-\text{i}\int_{0}^{t}\textrm{d}t'\hH(t')/\hbar}}{\mathbf{q}_\text{i}}$ can be given in terms of the classical action functional $\mathcal{S}[\mathbf{q},t']=\int_{0}^{t}\textrm{d}t'\mathcal{L}(\mathbf{q},t)$ as follows~\cite{Weiss2008}:
\begin{align}
K(\mathbf{q}_\text{f},\mathbf{q}_\text{i})=\int_{\mathbf{q}_\text{i}}^{\mathbf{q}_\text{f}}\mathcal{D}\mathbf{q} \exp\left\{\frac{\text{i}}{\hbar} \mathcal{S}[\mathbf{q},t']\right\},~\label{eq:K}
\end{align}
where the integration measure
\begin{align}
\int_{\mathbf{q}_\text{i}}^{\mathbf{q}_\text{f}}\mathcal{D}\mathbf{q}=\lim_{N\rightarrow\infty}\int_{-\infty}^{\infty}\frac{\textrm{d}^{N-1}\mathbf{q}}{c},
\end{align}
with $c$ being a normalization factor, goes along all possible paths $\mathbf{q}(t')$ between $\mathbf{q}_\text{i}=\mathbf{q}(0)$ and $\mathbf{q}_\text{f}=\mathbf{q}(t)$. Here, the temporal dependence of $\hH(t')$ is manifested by the explicit dependence on $t'$ in the classical Lagrangian $\mathcal{L}(\mathbf{q},t')=\mathcal{H}(\dot{\mathbf{q}},t')-\mathcal{V}(\mathbf{q},t')$, with $\mathcal{H}(\dot{\mathbf{q}},t')$ and $\mathcal{V}(\mathbf{q},t')$ being the kinetic-like and potential energy of the system, respectively. 

Suppose that the considered system is multipartite and one is just interested in the reduced dynamics of a subpartition with position and momentum operators $\hq$ and $\hp$, respectively, and the corresponding eigenstates $\ket{q}$ and $\ket{p}$. Moreover, suppose that the temporal dependence of $\hH(t)$ arises strictly from the free Hamiltonian of this main subsystem, denoted here by $\hH_\text{S}(t)$. By assuming that the state of the main subsystem is initially factorized from the rest, the path integral formalism allows one to express the reduced density operator in position representation, $\rho_{\text{S}}(q_\text{f},q'_\text{f})=\mel{q_\text{f}}{\hrh_\text{S}(t)}{q'_\text{f}}$, as
\begin{align}
\rho_{\text{S}}(q_{f},q'_{f})=&\int \textrm{d}q_{i}\textrm{d}q'_{i}\,\mathcal{J}(q_\text{f},q_\text{f}',q_\text{i},q_\text{i}')\rho_{\text{S}}(q_{i},q_{i}'),\nonumber\\
\mathcal{J}(q_\text{f},q_\text{f}',q_\text{i},q_\text{i}')=&\int_{q_{i}}^{q_{f}}\mathcal{D}q\int_{q'_{i}}^{q'_{f}}\mathcal{D}q'\,\text{e}^{\frac{\text{i}}{\hbar}\mathcal{S}_{\text{S}}[q,t]} \text{e}^{-\frac{\text{i}}{\hbar}\mathcal{S}_{\text{S}}[q',t]}F[q,q'],\label{eq:rhopath}
\end{align}
where classical action $\mathcal{S}_{\text{S}}[q,t]$ is associated with $\hH_\text{S}(t)$. All the dynamical effects of the secondary subsystems on the main one are encoded in the so-called influence functional $F[q,q']$, which equals unity in absence of interaction. 

In this work, one assumes that the secondary subsystems form a thermal bosonic bath and its interaction with the main subsystem is linear through the position coordinates as described by the Caldeira--Leggett model \cite{Caldeira1981}. If the main subsystem is a driven qubit as described in section~\ref{sec:model}, $\hH_\text{S}(t)=\hH_\text{S}+\hH_\text{d}(t)$, the Caldeira--Leggett model reduces to the Hamiltonian in equation~\eqref{eq:H1}, and the influence functional can be cast into the form $F[u,v]=\text{e}^{-\Phi[u,v]}$, where $\Phi[u,v]$ is a phase functional with real and imaginary parts \cite{Stockburger1999,Stockburger2004}
\begin{align}
\Phi_\text{r}[u]&=\int_{0}^{t} \textrm{d}t'\int_{0}^{t'}\textrm{d}t''u(t')u(t'')L_\textrm{r}(t'-t''),\label{eq:rphase}\\ 
\Phi_\text{i}[u,v]&=\int_{0}^{t} \textrm{d}t'\int_{0}^{t'}\textrm{d}t''u(t')v(t'')L_\textrm{i}(t'-t''),\label{eq:iphase}
\end{align}
where we defined new integration path variables $u=q-q'$ and $v=q+q'$.

Despite the exact expression for the influence functional $F[u,v]$, its calculation is nontrivial since it is nonlocal in time. For the choice of the spectral density $J(\omega)$ in equation~\eqref{eq:J2} and in the limit of high cutoff frequency ($\omega_\text{c} \rightarrow \infty$), such temporal nonlocality can be only attributed to the finite temperature of the bath. In this case, one can rewrite the phase functional as the sum of temporally nonlocal and temporally local phases, i.e., $\Phi[u,v]=\Phi_{\text{tnl}}[u]+\Phi_{\text{tl}}[u,v]$, with
\begin{align}
\Phi_{\text{tnl}}[u]&=\int_{0}^{t} \textrm{d}t'\int_{0}^{t'}\textrm{d}t''u(t')u(t'')L'_\text{r}(t'-t''),\label{eq:tnlphase}\\ 
\Phi_{\text{tl}}[u,v]&=\frac{\eta}{\hbar\beta}\int_{0}^{t} \textrm{d}t'u^2(t')+\text{i}\frac{\eta}{2} \int_{0}^{t} \textrm{d}t'u(t')\dot{v}(t').\label{eq:tlphase}
\end{align}
In equation~\eqref{eq:tnlphase} one has defined 
\begin{align}
L'_\text{r}(\tau)=&\int_{0}^{\infty}\frac{\textrm{d}\omega}{2\pi} J(\omega)[\coth(\hbar\beta \omega/2)-2/(\hbar\beta \omega)]\cos(\omega \tau) \label{eq:realmodL}
\end{align}
as the white noise deducted from the real part of $L(\tau)$. 
 
Here, one can establish a numerically exact correspondence of $F[u,v]$ with a temporally local averaged functional $\mathbb{E}\{F_{\xi}[u,v]\}$ arising from a classical stochastic process. This process is described by the real-valued classical random variable $\xi(t)$ with null mean and autocorrelation 
\begin{align}
\mathbb{E}\left[\xi(t')\xi(t'')\right]&=L'_\text{r}(t'-t''). \label{eq:corrfunc}
\end{align}
Placing equation~\eqref{eq:corrfunc} into equation~\eqref{eq:tnlphase} and
using a Hubbard--Stratonovich transformation \cite{Stratonovich1957,Hubbard1959}, one can write the influence functional $F[u,v]$ as 
\begin{align}
    \mathbb{E}\{F_{\xi}[u,v]\}=\text{e}^{-\Phi_{\text{tl}}[u,v]}\mathbb{E}\left[\text{e}^{\text{i}\int_{0}^{t}\textrm{d}t'u(t')\xi(t')}\right],   \label{eq:inffunc}
\end{align}
and consequently equation~\eqref{eq:rhopath} reduces to $\rho_{\text{S}}(q_\text{f},q'_\text{f})=\mathbb{E}[\rho_{\text{S},\xi}(q_\text{f},q'_\text{f})]$. Namely, the actual density operator $\rho_{\text{S}}(q_\text{f},q'_\text{f})$ can be regarded as the initial state $\rho_{\text{S}}(q_\text{i},q'_\text{i})$ evolving according to the stochastic influence functional $F_{\xi}[u,v]$ and averaged over a large number of noise trajectories. By returning to the operator representation and making the replacements $\hq\rightarrow \hsgx$, $\hp\rightarrow \omega_{\text{q}}\hsgy$, the evolution corresponding to a single realization of $\xi(t)$ is given by equation~\eqref{eq:SLED} of the main text. Such equation is deterministic and a single realization of $\xi(t)$ can be generated from an arbitrary Gaussian random variable (see~\ref{subsec:app1}). Interestingly, equation~\eqref{eq:SLED} has the form of the Caldeira--Leggett master equation~\cite{Heinz-PeterBreuer2007} with $-\xi(t)\hsgx$ added to the unitary part. Physically, such a term is responsible to account for the quantum fluctuations neglected in the classical treatment of the dissipative environment~\cite{Stockburger1999}. 

It has been shown~\cite{Stockburger2002,Stockburger2004} that the full stochastic unraveling of $\rho_{\text{S}}(q_\text{f},q'_\text{f})$ for an arbitrary spectral density $J(\omega)$ requires the inclusion of two complex-valued random variables, therefore making the numerical convergence slower than that of SLED. A thorough analysis involving such an extended method is out of the scope of this work.

\subsection{Noise generation in SLED} \label{subsec:app1}
    
Here, we describe the procedure for the generation of the stochastic noise $\xi(t)$ which appears in equation~\eqref{eq:SLED}. The initial point is to consider a Gaussian random variable $r(t)$, the autocorrelation function of which is given by
\begin{align}
    \mathbb{E}\left[r(t)r(t')\right]=\delta(t-t'). \label{eq:random}
\end{align}
Then one can define a real-valued convolution kernel $G(t)$ in such a way that the noise $\xi(t)$ is written as
\begin{align}
    \xi(t)=\int_{-\infty}^{\infty}\textrm{d}\tau\, G(t-\tau)r(\tau).\label{eq:xi1}
\end{align}
By using the definition~\eqref{eq:xi1} and the property~\eqref{eq:random}, the autocorrelation function of the noise $\xi(t)$ becomes
\begin{align}
    \mathbb{E}\left[\xi(t)\xi(t')\right]=\int_{-\infty}^{\infty}\textrm{d}\tau\, G(t-\tau)G(t'-\tau).\label{eq:corrfunc2}
\end{align}
The relation between equation~\eqref{eq:corrfunc2} and equation~\eqref{eq:corrfunc} can here be established by writing $G(t)$ and $L'_\text{r}(t)$ as the inverse associated to their Fourier transforms 
\begin{align}
\tilde{G}(\omega)&=\int_{-\infty}^{\infty}\textrm{d}t\, \text{e}^{-\text{i}\omega t}G(t),\label{eq:Gw}\\
\tilde{L}'_\text{r}(\omega)&=\int_{-\infty}^{\infty}\textrm{d}t\, \text{e}^{-\text{i}\omega t}L'_\text{r}(t),\label{eq:modrLw}
\end{align}
that is to say, 
\begin{align}
G(t)&=\frac{1}{2\pi}\int_{-\infty}^{\infty}\textrm{d}\omega\, \text{e}^{\text{i}\omega t}\tilde{G}(\omega),\label{eq:Gt}\\
L'_\text{r}(t)&=\frac{1}{2\pi}\int_{-\infty}^{\infty}\textrm{d}\omega\, \text{e}^{\text{i}\omega t}\tilde{L}'_\text{r}(\omega).\label{eq:modrLt}
\end{align}
Equating equation~\eqref{eq:corrfunc2} with equation~\eqref{eq:corrfunc} allows one to obtain
\begin{align}
    \tilde{G}(\omega)=\sqrt{\tilde{L}'_\text{r}(\omega)}.\label{eq:Gw2}
\end{align}
For an odd spectral density $J(\omega)$, as the one defined in equation~\eqref{eq:J2}, the Fourier transform of $L'_\text{r}(t)$ acquires the form
\begin{align}
    \tilde{L}'_\text{r}(\omega)=J(\omega)[\coth(\hbar\beta\omega/2)-2/(\hbar\beta\omega)]/2. \label{eq:modrLw2}
\end{align}
Finally, by denoting the Fourier transform of $r(t)$ as $\tilde{r}(\omega)$, equation~\eqref{eq:xi1} can be rewritten as 
\begin{align}
    \xi(t)=\frac{1}{2\pi}\int_{-\infty}^{\infty}\textrm{d}\omega\, \text{e}^{\text{i}\omega t}\tilde{G}(\omega)\tilde{r}(\omega) \label{eq:xi2}.
\end{align}
Therefore, $\xi(t)$ can be regarded as the inverse transform of $\tilde{G}(\omega)\tilde{r}(\omega)$, with $\tilde{G}(\omega)$ being obtained through equation~\eqref{eq:Gw2} and $\tilde{r}(\omega)$ being generated from a Gaussian random variable $r(t)$. In this work, we have used Python built-in functions for generating $r(t)$ and for calculating the Fourier and inverse transforms.

\section{Phenomenological description of the experimental setup} \label{sec:app2}

In this section, we phenomenologically describe a cQED setup where pump-probe measurements may be carried out in the presence of tunable dissipation. For simplicity, we consider a transmon qubit with the transition frequency $\omega_{\text{q}}$ between its two lowest levels $\ket{i}$ $(i=0,1)$, which is capacitively coupled to microwave drive and probe lines characterized by a time dependent voltage \cite{Krantz2019}. In addition, the qubit is capacitively coupled to a tunable resistor at inverse temperature $\beta$, which can be implemented either through a quantum circuit refrigerator or a heat sink \cite{Tuorila2019}. These features may be modeled by the Hamiltonian~\eqref{eq:H1} of the main text with the form of $f(t)$ given by equation~\eqref{eq:f2}.

The readout of the qubit state can be achieved by the measurement of the transmitted field through a resonator coupled dispersively
to the transmon. Namely, for a resonator with angular frequency $\omega_{\text{{\rm r}}}$,
photon decay rate $\kappa$, and annihilation operator
$\ha$, coupled linearly to the transmon via a Jaynes--Cummings
interaction $\hH_{\text{JC}}=\hbar g\left(\kb{1}{0}\ha+\kb{0}{1}\hdgg a\right)$,
the dispersive condition ($g/\Delta_{{\rm qr}}\ll1$, $\Delta_{{\rm qr}}=\omega_{{\rm q}}-\omega_{{\rm r}}$)
assures that essentially no energy is exchanged between them; only frequency shifts
are induced \cite{Teixeira2019}. The phenomenological inclusion of the resonator in the dispersive regime thus produces the effective system Hamiltonian
\begin{align}
\hH_{\text{qr}}(t)=&-\frac{\hbar\tilde{\omega}_{\text{q}}}{2}\hsgz+\hbar f(t)\hsgx+\hbar\left(\omega_{\text{r}}+\chi\hsgz\right)\hdgg a\ha\nonumber \\
{}&+\frac{\hbar \Omega_{\text{m}}}{2}\left(\hdgg a\text{e}^{-\text{i}\omega_{\text{m}}t}+\ha \text{e}^{\text{i}\omega_{\text{m}}t}\right),\label{eq:Heff}
\end{align}
where $\tilde{\omega}_{\text{q}}=\omega_{\text{q}}+\chi$ and $\chi=g^{2}/\Delta_{\text{r}}$
is the frequency shift induced by the dispersive interaction. Note that we also
consider a weak measurement drive of frequency $\omega_{\text{m}}$
and amplitude $\Omega_{\text{m}}$ applied to the input port of the resonator. 

Here, we make the assumption that the open dynamics of the qubit-resonator system is provided by local quantum environments. This is augmented by the fact that the incoherent dynamics of the whole system is caused by independent sources. For simplicity, we neglect intrinsic uncontrollable dephasing and dissipation rates of the qubit in the treatment by assuming that they are much smaller than the dissipation rate $\gamma$ produced by its coupling to the tunable resistor, the latter thus being the main source of dissipation for the qubit within the considered times scales. On the other hand, the dissipative
dynamics of the resonator is caused by its finite quality factor so that it behaves as a lossy cavity where photons can leak out incoherently. These features can be represented by a master equation of the form
\begin{align}
\dert{\hrh_{\text{qr}}(t)}=-\frac{\text{i}}{\hbar}\left[\hH_{\text{qr}}(t),\hrh_{\text{qr}}(t)\right]+\mathcal{L}_{\gamma}\left[\hrh_{\text{qr}}(t)\right]+\mathcal{L}_{\kappa}\left[\hrh_{\text{qr}}(t)\right]\label{eq:meqSLED-Lind}
\end{align}
where the commutator describes the unitary dynamics determined by the Hamiltonian in equation~\eqref{eq:Heff}, and $\mathcal{L}_{\gamma/\kappa}\left[\hrh_{\text{qr}}(t)\right]$ describe the non-unitary dynamics promoted by the tunable resistor and the lossy resonator, respectively. Hence, $\mathcal{L}_{\gamma/\kappa}\left[\hrh_{\text{qr}}(t)\right]$ contains only operators either in the qubit or the resonator subspace and already includes all possible bath-induced energy shifts and dissipative effects. 

Based on the arguments presented above, equation~(\ref{eq:meqSLED-Lind}) allows one to separate the non-unitary effects produced by each local bath and describe the mutually induced frequency shifts in the qubit-resonator system by the effective Hamiltonian $\hH_{\text{qr}}(t)$. While the average number of photons in the resonator contributes to the shift of the qubit frequency, the qubit-dependent
frequency of the resonator changes the behavior of the transmitted
field providing an indirect measurement of the driven qubit spectrum as a function of the probe frequency $\omega_{\text{p}}$. In order to visualize such a phenomenon, we study the temporal evolution of the expectation value of $\ha$ in a frame rotating at $\omega_{\text{m}}$, $a(t)=\Tr[\ha \text{e}^{\text{i}\omega_{{\rm m}}t\hdgg a\ha}\hrh_{\text{qr}}(t)\text{e}^{-\text{i}\omega_{{\rm m}}t\hdgg a\ha}]$. First, we assume that $\mathcal{L}_{\kappa}\left[\hrh_{\text{qr}}(t)\right]$ is phenomenologically described in the Lindblad form
\begin{align}
    \mathcal{L}_{\kappa}\left[\hrh_{\text{qr}}(t)\right]=\kappa\left[\ha\hrh_{\text{qr}}(t)\hdgg a-\frac{1}{2}\left\{ \hdgg a\ha,\hrh_{\text{qr}}(t)\right\} \right], \label{eq:dissip}
\end{align}
where for shortness of notation we omitted the energy shift term, assuming that it is already incorporated to the definition of $\omega_{\text{r}}$. 
Using equation~\eqref{eq:meqSLED-Lind}, we can write the dynamical equation for the expectation value of the annihilation operator as
\begin{equation}
\dot{a}(t)=-\text{i}\frac{\Omega_{{\rm m}}}{2}-\left(i\Delta_{{\rm rm}}+\frac{\kappa}{2}\right)a(t)-\text{i}\chi a_z(t),\label{eq:aeq1}
\end{equation}
where we have defined $\Delta_{{\rm rm}}=\omega_{{\rm r}}-\omega_{{\rm m}}$ and $a_z(t)=\Tr[\hsgz\ha \text{e}^{\text{i}\omega_{{\rm m}}t\hdgg a\ha}\hrh_{\text{qr}}(t)\text{e}^{-\text{i}\omega_{{\rm m}}t\hdgg a\ha}]$. In the semiclassical approximation \cite{Tuorila2009}, one can neglect the influence of qubit-resonator entanglement on the temporal evolution of $a_z(t)$, in such a way that $a_z(t)\approx a(t)\sigma_z(t)$, with $\sigma_z(t)=\Tr[\hsgz\hrh_{\text{qr}}(t)]$. Consequently, equation~\eqref{eq:aeq1} can be rewritten as
\begin{equation}
\dot{a}(t)=-\text{i}\frac{\Omega_{{\rm m}}}{2}-\left[\text{i}\Delta_{{\rm rm}}+\text{i}\chi \sigma_z(t)+\frac{\kappa}{2}\right]a(t).\label{eq:aeq2}
\end{equation}
Here, we assume that the measurement field
$\Omega_{\text{m}}$ is turned on at a sufficiently long time after the initial transient dynamics of the dissipative driven qubit. Consequently, provided that the probe is much weaker than the drive, i.e., $\Omega_{\text{p}}\ll\Omega_{\text{d}}$, we can write the solution to equation~\eqref{eq:aeq2} as
\begin{align}
a(t)=&\left[a(0)-\text{i}\Omega_{\text{m}}\left(\frac{\text{e}^{\left[\text{i}\left(\Delta_{\text{rm}}+\chi\bar{\sigma}_z\right)+\kappa/2\right]t}-1}{\kappa+2\text{i}\left(\Delta_{\text{rm}}+\chi\bar{\sigma}_z\right)}\right)\right] \text{e}^{-\left[\text{i}\left(\Delta_{\text{rm}}+\chi\bar{\sigma}_z\right)+\kappa/2\right]t},\label{eq:asol2}
\end{align}
where $\bar{\sigma}_z$ is the temporal average of $\sigma_z(t)$. For the choice of $t=n_{\text{p}}t_{\text{p}}$ as in section~\ref{sec:mollow}, $\bar{\sigma}_z$ may be written as in equation~\eqref{eq:tavgz}.

In our approach, we solve the dissipative dynamics of the driven qubit and feed the solution of $a(t)$ with the values of $\bar{\sigma}_z$. As explained in the main text, in this work $\sigma_z(t)$ is obtained either from the Lindblad master equation~\eqref{eq:drho4} or by the average solution of the SLED in equation~\eqref{eq:SLED}. Regardless on the method of solution of the qubit dynamics, equation~\eqref{eq:asol2} clearly shows that $a(t)$ contains information about
the qubit population and, therefore, contains information about its spectrum. One can access the features of the spectrum, for instance, through the amplitude, phase, or Fourier transform of the transmitted field. Defining the field quadratures as $I(t)=\text{Re}[a(t)]$ and $Q(t)=\text{Im}[a(t)]$, the amplitude $A(t)$ and phase $\phi(t)$ of the transmitted signal can be expressed as \cite{Krantz2019}
\begin{align}
    A(t)=\sqrt{I^{2}(t)+Q^{2}(t)},\,\,\,\phi(t)=\text{Arg}[a(t)]. \label{eq:AandPhi}
\end{align}
Finally, by setting $\Delta_{{\rm rm}}=0$, assuming $\kappa t/2=\kappa n_{\text{p}}t_{\text{p}}/2\gg1$, and using the definition for $A(t)$ in equation~\eqref{eq:AandPhi}, one finds the transmitted field amplitude
\begin{align}
A = \frac{\Omega_{\text{m}}}{\kappa}\frac{1}{\sqrt{1+\left(\frac{2\chi \bar{\sigma}_z}{\kappa}\right)^2}},
\end{align}
as defined in equation~\eqref{eq:Amp}.
	
	\footnotesize
    \bibliography{refs}

\providecommand{\newblock}{}
\begin{thebibliography}{100}
\expandafter\ifx\csname url\endcsname\relax
  \def\url#1{{\tt #1}}\fi
\expandafter\ifx\csname urlprefix\endcsname\relax\def\urlprefix{URL }\fi
\providecommand{\eprint}[2][]{\url{#2}}

\bibitem{CrispinGardiner2015}
Gardiner C and Zoller P 2015 {\em The Quantum World of Ultra-Cold Atoms and
  Light II\/} (London: Imperial College Press)

\bibitem{Weiss2008}
Weiss U 2008 {\em Quantum Dissipative Systems, 3rd ed.\/} (Singapore: World
  Scientific)

\bibitem{Cirac1995}
Cirac J~I and Zoller P 1995 {\em Phys. Rev. Lett.\/}
  \href{http://dx.doi.org/10.1103/physrevlett.74.4091}{74, 4091}

\bibitem{Blais2004}
Blais A, Huang R~S, Wallraff A, Girvin S~M and Schoelkopf R~J 2004 {\em Phys.
  Rev. A\/} \href{http://dx.doi.org/10.1103/physreva.69.062320}{69, 062320}

\bibitem{Nakamura1999}
Nakamura Y, Pashkin Y~A and Tsai J~S 1999 {\em Nature\/}
  \href{http://dx.doi.org/10.1038/19718}{398, 786}

\bibitem{Cole2001}
Cole B~E, Williams J~B, King B~T, Sherwin M~S and Stanley C~R 2001 {\em
  Nature\/} \href{http://dx.doi.org/10.1038/35065032}{410, 60}

\bibitem{Letchumanan2004}
Letchumanan V, Gill P, Riis E and Sinclair A~G 2004 {\em Phys. Rev. A\/}
  \href{http://dx.doi.org/10.1103/physreva.70.033419}{70, 033419}

\bibitem{Eckel2009}
Eckel J, Reina J~H and Thorwart M 2009 {\em New J. Phys.\/}
  \href{http://dx.doi.org/10.1088/1367-2630/11/8/085001}{11, 085001}

\bibitem{Medina2019}
Medina I and Semi{\~{a}}o F~L 2019 {\em Phys. Rev. A\/}
  \href{http://dx.doi.org/10.1103/physreva.100.012103}{100, 012103}

\bibitem{Gelman2005}
Gelman D and Kosloff R 2005 {\em J. Chem. Phys.\/}
  \href{http://dx.doi.org/10.1063/1.2136155}{123, 234506}

\bibitem{Prokhorenko2006}
Prokhorenko V~I 2006 {\em Science\/}
  \href{http://dx.doi.org/10.1126/science.1130747}{313, 1257}

\bibitem{Golubev2015}
Golubev N~V and Kuleff A~I 2015 {\em Phys. Rev. A\/}
  \href{http://dx.doi.org/10.1103/physreva.91.051401}{91, 051401}

\bibitem{Wilson2010}
Wilson C~M, Johansson G, Duty T, Persson F, Sandberg M and Delsing P 2010 {\em
  Phys. Rev. B\/} \href{http://dx.doi.org/10.1103/PhysRevB.81.024520}{81,
  024520}

\bibitem{Grifoni1998}
Grifoni M and H\"anggi P 1998 {\em Phys. Rep.\/}
  \href{http://dx.doi.org/10.1016/s0370-1573(98)00022-2}{304, 229}

\bibitem{Silveri2017}
Silveri M~P, Tuorila J~A, Thuneberg E~V and Paraoanu G~S 2017 {\em Rep. Prog.
  Phys.\/} \href{http://dx.doi.org/10.1088/1361-6633/aa5170}{80, 056002}

\bibitem{Nakamura2001}
Nakamura Y, Pashkin Y~A and Tsai J~S 2001 {\em Phys. Rev. Lett.\/}
  \href{http://dx.doi.org/10.1103/PhysRevLett.87.246601}{87, 246601}

\bibitem{Tuorila2010}
Tuorila J, Silveri M, Sillanp\"a\"a M, Thuneberg E, Makhlin Y and Hakonen P
  2010 {\em Phys. Rev. Lett.\/}
  \href{http://dx.doi.org/10.1103/PhysRevLett.105.257003}{105, 257003}

\bibitem{Deng2015}
Deng C, Orgiazzi J~L, Shen F, Ashhab S and Lupascu A 2015 {\em Phys. Rev.
  Lett.\/} \href{http://dx.doi.org/10.1103/PhysRevLett.115.133601}{115, 133601}

\bibitem{Grossmann1991}
Grossmann F, Dittrich T, Jung P and H\"anggi P 1991 {\em Phys. Rev. Lett.\/}
  \href{http://dx.doi.org/10.1103/PhysRevLett.67.516}{67, 516}

\bibitem{Dakhnovskii1993}
Dakhnovskii Y and Metiu H 1993 {\em Phys. Rev. A\/}
  \href{http://dx.doi.org/10.1103/PhysRevA.48.2342}{48, 2342}

\bibitem{Shevchenko2010}
Shevchenko S~N, Ashhab S and Nori F 2010 {\em Phys. Rep.\/}
  \href{http://dx.doi.org/https://doi.org/10.1016/j.physrep.2010.03.002}{492,
  1}

\bibitem{Oliver2005}
Oliver W~D, Yu Y, Lee J~C, Berggren K~K, Levitov L~S and Orlando T~P 2005 {\em
  Science\/} \href{http://dx.doi.org/10.1126/science.1119678}{310, 1653}

\bibitem{Sillanpaa2006}
Sillanp\"a\"a M, Lehtinen T, Paila A, Makhlin Y and Hakonen P 2006 {\em Phys.
  Rev. Lett.\/} \href{http://dx.doi.org/10.1103/PhysRevLett.96.187002}{96,
  187002}

\bibitem{Silveri2015}
Silveri M~P, Kumar K~S, Tuorila J, Li J, Veps\"al\"ainen A, Thuneberg E~V and
  Paraoanu G~S 2015 {\em New J. Phys.\/}
  \href{http://dx.doi.org/10.1088/1367-2630/17/4/043058}{17, 043058}

\bibitem{Tuorila2013}
Tuorila J, Silveri M, Sillanp\"a\"a M, Thuneberg E, Makhlin Y and Hakonen P
  2013 {\em Supercond. Sci. Technol.\/}
  \href{http://dx.doi.org/10.1088/0953-2048/26/12/124001}{26, 124001}

\bibitem{Mollow1969}
Mollow B~R 1969 {\em Phys. Rev.\/}
  \href{http://dx.doi.org/10.1103/physrev.188.1969}{188, 1969}

\bibitem{Oliver1971}
Oliver G, Ressayre E and Tallet A 1971 {\em Lett. Nuovo Cimento\/}
  \href{http://dx.doi.org/10.1007/bf02789642}{2, 777}

\bibitem{Carmichael1976}
Carmichael H~J and Walls D~F 1976 {\em J. Phys. B: At. Mol. Phys.\/}
  \href{http://dx.doi.org/10.1088/0022-3700/9/8/007}{9, 1199}

\bibitem{Cohen-Tannoudji1977}
Cohen-Tannoudji C and Reynaud S 1977 {\em J. Phys. B: At. Mol. Phys.\/}
  \href{http://dx.doi.org/10.1088/0022-3700/10/3/005}{10, 345}

\bibitem{Schuda1974}
Schuda F, Stroud C~R and Hercher M 1974 {\em J. Phys. B: At. Mol. Phys.\/}
  \href{http://dx.doi.org/10.1088/0022-3700/7/7/002}{7, L198}

\bibitem{Wrigge2007}
Wrigge G, Gerhardt I, Hwang J, Zumofen G and Sandoghdar V 2007 {\em Nat.
  Phys.\/} \href{http://dx.doi.org/10.1038/nphys812}{4, 60}

\bibitem{Baur2009}
Baur M, Filipp S, Bianchetti R, Fink J~M, G\"oppl M, Steffen L, Leek P~J, Blais
  A and Wallraff A 2009 {\em Phys. Rev. Lett.\/}
  \href{http://dx.doi.org/10.1103/physrevlett.102.243602}{102, 243602}

\bibitem{Astafiev2010}
Astafiev O, Zagoskin A~M, Abdumalikov A~A, Pashkin Y~A, Yamamoto T, Inomata K,
  Nakamura Y and Tsai J~S 2010 {\em Science\/}
  \href{http://dx.doi.org/10.1126/science.1181918}{327, 840}

\bibitem{Ulhaq2013}
Ulhaq A, Weiler S, Roy C, Ulrich S~M, Jetter M, Hughes S and Michler P 2013
  {\em Opt. Express\/} \href{http://dx.doi.org/10.1364/oe.21.004382}{21, 4382}

\bibitem{Unsleber2015}
Unsleber S, Maier S, McCutcheon D~P~S, He Y~M, Dambach M, Gschrey M, Gregersen
  N, M{\o}rk J, Reitzenstein S, H\"ofling S, Schneider C and Kamp M 2015 {\em
  Optica\/} \href{http://dx.doi.org/10.1364/optica.2.001072}{2, 1072}

\bibitem{Pigeau2015}
Pigeau B, Rohr S, de~L{\'{e}}pinay L~M, Gloppe A, Jacques V and Arcizet O 2015
  {\em Nat. Commun.\/} \href{http://dx.doi.org/10.1038/ncomms9603}{6, 8603}

\bibitem{Lagoudakis2017}
Lagoudakis K~G, Fischer K~A, Sarmiento T, McMahon P~L, Radulaski M, Zhang J~L,
  Kelaita Y, Dory C, M\"uller K and Vu{\v{c}}kovi{\'{c}} J 2017 {\em Phys. Rev.
  Lett.\/} \href{http://dx.doi.org/10.1103/PhysRevLett.118.013602}{118, 013602}

\bibitem{Ortiz-Gutierrez2019}
Ortiz-Guti{\'{e}}rrez L, Teixeira R~C, Eloy A, da~Silva D~F, Kaiser R,
  Bachelard R and Fouch{\'{e}} M 2019 {\em New J. Phys.\/}
  \href{http://dx.doi.org/10.1088/1367-2630/ab3ca9}{21, 093019}

\bibitem{Carmichael2013}
Carmichael H 2013 {\em An Open Systems Approach to Quantum Optics\/} (Berlin:
  Springer)

\bibitem{Heinz-PeterBreuer2007}
Breuer H~P and Petruccione F 2007 {\em The Theory of Open Quantum Systems\/}
  (Oxford: Oxford University Press)

\bibitem{CrispinGardiner2004}
Gardiner C and Zoller P 2004 {\em Quantum Noise\/} (Berlin: Springer)

\bibitem{Lindblad1976}
Lindblad G 1976 {\em Commun. Math. Phys.\/}
  \href{http://dx.doi.org/10.1007/bf01608499}{48, 119}

\bibitem{Gorini1976}
Gorini V, Kossakowski A and Sudarshan E~C~G 1976 {\em J. Math. Phys.\/}
  \href{http://dx.doi.org/10.1063/1.522979}{17, 821}

\bibitem{Pekola2010}
Pekola J~P, Brosco V, M\"ott\"onen M, Solinas P and Shnirman A 2010 {\em Phys.
  Rev. Lett.\/} \href{http://dx.doi.org/10.1103/physrevlett.105.030401}{105,
  030401}

\bibitem{Salmilehto2010}
Salmilehto J, Solinas P, Ankerhold J and M\"ott\"onen M 2010 {\em Phys. Rev.
  A\/} \href{http://dx.doi.org/10.1103/physreva.82.062112}{82, 062112}

\bibitem{Salmilehto2011}
Salmilehto J and M\"ott\"onen M 2011 {\em Phys. Rev. B\/}
  \href{http://dx.doi.org/10.1103/physrevb.84.174507}{84, 174507}

\bibitem{Albash2012}
Albash T, Boixo S, Lidar D~A and Zanardi P 2012 {\em New J. Phys.\/}
  \href{http://dx.doi.org/10.1088/1367-2630/14/12/123016}{14, 123016}

\bibitem{Xu2014}
Xu C, Poudel A and Vavilov M~G 2014 {\em Phys. Rev. A\/}
  \href{http://dx.doi.org/10.1103/PhysRevA.89.052102}{89, 052102}

\bibitem{Blumel1991}
Bl\"umel R, Buchleitner A, Graham R, Sirko L, Smilansky U and Walther H 1991
  {\em Phys. Rev. A\/} \href{http://dx.doi.org/10.1103/PhysRevA.44.4521}{44,
  4521}

\bibitem{Dittrich1993}
Dittrich T, Oelschl\"agel B and H\"anggi P 1993 {\em Europhys. Lett.\/}
  \href{http://dx.doi.org/10.1209/0295-5075/22/1/002}{22, 5}

\bibitem{Hausinger2010}
Hausinger J and Grifoni M 2010 {\em Phys. Rev. A\/}
  \href{http://dx.doi.org/10.1103/PhysRevA.81.022117}{81, 022117}

\bibitem{Engelhardt2019}
Engelhardt G, Platero G and Cao J 2019 {\em Phys. Rev. Lett.\/}
  \href{http://dx.doi.org/10.1103/PhysRevLett.123.120602}{123, 120602}

\bibitem{Hartmann2000}
Hartmann L, Goychuk I, Grifoni M and H\"anggi P 2000 {\em Phys. Rev. E\/}
  \href{http://dx.doi.org/10.1103/PhysRevE.61.R4687}{61, R4687}

\bibitem{Ikeda2020}
Ikeda T~N and Sato M 2020 {\em Sci. Adv.\/}
  \href{http://dx.doi.org/10.1126/sciadv.abb4019}{6, eabb4019}

\bibitem{Leggett1987}
Leggett A~J, Chakravarty S, Dorsey A~T, Fisher M~P~A, Garg A and Zwerger W 1987
  {\em Rev. Mod. Phys.\/} \href{http://dx.doi.org/10.1103/RevModPhys.59.1}{59,
  1}

\bibitem{Leggett1995}
Leggett A~J, Chakravarty S, Dorsey A~T, Fisher M~P~A, Garg A and Zwerger W 1995
  {\em Rev. Mod. Phys.\/}
  \href{http://dx.doi.org/10.1103/RevModPhys.67.725}{67, 725}

\bibitem{Grifoni1995}
Grifoni M, Sassetti M, H\"anggi P and Weiss U 1995 {\em Phys. Rev. E\/}
  \href{http://dx.doi.org/10.1103/PhysRevE.52.3596}{52, 3596}

\bibitem{Magazzu2018b}
Magazz\`u L, Forn-D\'iaz P, Belyansky R, Orgiazzi J~L, Yurtalan M~A, Otto M~R,
  Lupascu A, Wilson C~M and Grifoni M 2018 {\em Nat. Commun.\/}
  \href{http://dx.doi.org/10.1038/s41467-018-03626-w}{9, 1403}

\bibitem{McCutcheon2011}
McCutcheon D~P~S, Dattani N~S, Gauger E~M, Lovett B~W and Nazir A 2011 {\em
  Phys. Rev. B\/} \href{http://dx.doi.org/10.1103/physrevb.84.081305}{84,
  081305}

\bibitem{McCutcheon2013}
McCutcheon D~P~S and Nazir A 2013 {\em Phys. Rev. Lett.\/}
  \href{http://dx.doi.org/10.1103/physrevlett.110.217401}{110, 217401}

\bibitem{Restrepo2016}
Restrepo S, Cerrillo J, Bastidas V~M, Angelakis D~G and Brandes T 2016 {\em
  Phys. Rev. Lett.\/}
  \href{http://dx.doi.org/10.1103/PhysRevLett.117.250401}{117, 250401}

\bibitem{Prior2010}
Prior J, Chin A~W, Huelga S~F and Plenio M~B 2010 {\em Phys. Rev. Lett.\/}
  \href{http://dx.doi.org/10.1103/physrevlett.105.050404}{105, 050404}

\bibitem{Chin2010}
Chin A~W, Rivas {\'{A}}, Huelga S~F and Plenio M~B 2010 {\em J. Math. Phys\/}
  \href{http://dx.doi.org/10.1063/1.3490188}{51, 092109}

\bibitem{Makarov1994}
Makarov D~E and Makri N 1994 {\em Chem. Phys. Lett.\/}
  \href{http://dx.doi.org/https://doi.org/10.1016/0009-2614(94)00275-4}{221,
  482}

\bibitem{Makri1995}
Makri N 1995 {\em J. Math. Phys.\/}
  \href{http://dx.doi.org/10.1063/1.531046}{36, 2430}

\bibitem{Makri1995b}
Makri N and Makarov D~E 1995 {\em J. Chem. Phys.\/}
  \href{http://dx.doi.org/10.1063/1.469509}{102, 4611}

\bibitem{Tanimura1989}
Tanimura Y and Kubo R 1989 {\em J. Phys. Soc. Jpn.\/}
  \href{http://dx.doi.org/10.1143/jpsj.58.101}{58, 101}

\bibitem{Tanimura2020}
Tanimura Y 2020 {\em J. Chem. Phys.\/}
  \href{http://dx.doi.org/10.1063/5.0011599}{153, 020901}

\bibitem{Cangemi2019}
Cangemi L~M, Cataudella V, Sassetti M and Filippis G~D 2019 {\em Phys. Rev.
  B\/} \href{http://dx.doi.org/10.1103/physrevb.100.014301}{100, 014301}

\bibitem{Strathearn2018}
Strathearn A, Kirton P, Kilda D, Keeling J and Lovett B~W 2018 {\em Nat.
  Commun.\/} \href{http://dx.doi.org/10.1038/s41467-018-05617-3}{9, 1}

\bibitem{Cygorek2021}
Cygorek M, Cosacchi M, Vagov A, Axt V~M, Lovett B~W, Keeling J and Gauger E~M
  2021 Numerically exact open quantum systems simulations for arbitrary
  environments using automated compression of environments (\textit{Preprint}
  \eprint{2101.01653})

\bibitem{Stockburger1999}
Stockburger J~T and Mak C~H 1999 {\em J. Chem. Phys.\/}
  \href{http://dx.doi.org/10.1063/1.478396}{110, 4983}

\bibitem{Stockburger1999b}
Stockburger J~T 1999 {\em Phys. Rev. E\/}
  \href{http://dx.doi.org/10.1103/PhysRevE.59.R4709}{59, R4709}

\bibitem{Schmidt2011}
Schmidt R, Negretti A, Ankerhold J, Calarco T and Stockburger J~T 2011 {\em
  Phys. Rev. Lett.\/}
  \href{http://dx.doi.org/10.1103/physrevlett.107.130404}{107, 130404}

\bibitem{Schmidt2013}
Schmidt R, Stockburger J~T and Ankerhold J 2013 {\em Phys. Rev. A\/}
  \href{http://dx.doi.org/10.1103/physreva.88.052321}{88, 052321}

\bibitem{Tuorila2019}
Tuorila J, Stockburger J, Ala-Nissil\"a T, Ankerhold J and M\"ott\"onen M 2019
  {\em Phys. Rev. Research\/}
  \href{http://dx.doi.org/10.1103/physrevresearch.1.013004}{1, 013004}

\bibitem{Vadimov2021}
Vadimov V, Tuorila J, Orell T, Stockburger J, Ala-Nissila T, Ankerhold J and
  Möttönen M 2021 {\em Phys. Rev. B\/}
  \href{http://dx.doi.org/10.1103/physrevb.103.214308}{103, 214308}

\bibitem{Viitanen2021}
Viitanen A, Silveri M, Jenei M, Sevriuk V, Tan K~Y, Partanen M, Goetz J,
  Gr\"onberg L, Lahtinen V and M\"ott\"onen M 2021 {\em Phys. Rev. Research\/}
  \href{http://dx.doi.org/10.1103/physrevresearch.3.033126}{3, 033126}

\bibitem{Partanen2019}
Partanen M, Goetz J, Tan K~Y, Kohvakka K, Sevriuk V, Lake R~E, Kokkoniemi R,
  Ikonen J, Hazra D, M\"akinen A, Hyypp\"a E, Gr\"onberg L, Vesterinen V,
  Silveri M and M\"ott\"onen M 2019 {\em Phys. Rev. B\/}
  \href{http://dx.doi.org/10.1103/physrevb.100.134505}{100, 134505}

\bibitem{Sevriuk2019}
Sevriuk V~A, Tan K~Y, Hyypp\"a E, Silveri M, Partanen M, Jenei M, Masuda S,
  Goetz J, Vesterinen V, Gr\"onberg L and M\"ott\"onen M 2019 {\em Appl. Phys.
  Lett.\/} \href{http://dx.doi.org/10.1063/1.5116659}{115, 082601}

\bibitem{Silveri2019}
Silveri M, Masuda S, Sevriuk V, Tan K~Y, Jenei M, Hyypp\"a E, Hassler F,
  Partanen M, Goetz J, Lake R~E, Gr\"onberg L and M\"ott\"onen M 2019 {\em Nat.
  Phys.\/} \href{http://dx.doi.org/10.1038/s41567-019-0449-0}{15, 533}

\bibitem{Partanen2018}
Partanen M, Tan K~Y, Masuda S, Govenius J, Lake R~E, Jenei M, Gr\"onberg L,
  Hassel J, Simbierowicz S, Vesterinen V, Tuorila J, Ala-Nissil\"a T and
  M\"ott\"onen M 2018 {\em Sci. Rep.\/}
  \href{http://dx.doi.org/10.1038/s41598-018-24449-1}{8, 6325}

\bibitem{Tan2017}
Tan K~Y, Partanen M, Lake R~E, Govenius J, Masuda S and M\"ott\"onen M 2017
  {\em Nat. Commun.\/} \href{http://dx.doi.org/10.1038/ncomms15189}{8, 15189}

\bibitem{Harrington2019}
Harrington P~M, Naghiloo M, Tan D and Murch K~W 2019 {\em Phys. Rev. A\/}
  \href{http://dx.doi.org/10.1103/physreva.99.052126}{99, 052126}

\bibitem{Martinez2019}
Mart{\'{\i}}nez J~P, L{\'{e}}ger S, Gheeraert N, Dassonneville R, Planat L,
  Foroughi F, Krupko Y, Buisson O, Naud C, Hasch-Guichard W, Florens S, Snyman
  I and Roch N 2019 {\em npj Quantum Inf.\/}
  \href{http://dx.doi.org/10.1038/s41534-018-0104-0}{5, 19}

\bibitem{Ronzani2018}
Ronzani A, Karimi B, Senior J, Chang Y~C, Peltonen J~T, Chen C and Pekola J~P
  2018 {\em Nat. Phys.\/}
  \href{http://dx.doi.org/10.1038/s41567-018-0199-4}{14, 991}

\bibitem{Kimchi-Schwartz2016}
Kimchi-Schwartz M, Martin L, Flurin E, Aron C, Kulkarni M, Tureci H and Siddiqi
  I 2016 {\em Phys. Rev. Lett.\/}
  \href{http://dx.doi.org/10.1103/physrevlett.116.240503}{116, 240503}

\bibitem{Murch2012}
Murch K~W, Vool U, Zhou D, Weber S~J, Girvin S~M and Siddiqi I 2012 {\em Phys.
  Rev. Lett.\/} \href{http://dx.doi.org/10.1103/physrevlett.109.183602}{109,
  183602}

\bibitem{Orth2013}
Orth P~P, Imambekov A and Le~Hur K 2013 {\em Phys. Rev. B\/}
  \href{http://dx.doi.org/10.1103/PhysRevB.87.014305}{87, 014305}

\bibitem{Vega2010}
de~Vega I, Ba{\~{n}}uls M~C and P{\'{e}}rez A 2010 {\em New J. Phys.\/}
  \href{http://dx.doi.org/10.1088/1367-2630/12/12/123010}{12, 123010}

\bibitem{Thingna2012}
Thingna J, Wang J~S and Hänggi P 2012 {\em J. Chem. Phys.\/}
  \href{http://dx.doi.org/10.1063/1.4718706}{136, 194110}

\bibitem{Dutra1994}
Dutra S~M, Knight P~L and Moya-Cessa H 1994 {\em Phys. Rev. A\/}
  \href{http://dx.doi.org/10.1103/PhysRevA.49.1993}{49, 1993}

\bibitem{Salmilehto2014}
Salmilehto J, Solinas P and M\"ott\"onen M 2014 {\em Phys. Rev. E\/}
  \href{http://dx.doi.org/10.1103/physreve.89.052128}{89, 052128}

\bibitem{Ikonen2017}
Ikonen J, Salmilehto J and M\"ott\"onen M 2017 {\em npj Quantum Inf.\/}
  \href{http://dx.doi.org/10.1038/s41534-017-0015-5}{3, 17}

\bibitem{Purcell1946}
Purcell E~M, Torrey H~C and Pound R~V 1946 {\em Phys. Rev.\/}
  \href{http://dx.doi.org/10.1103/physrev.69.37}{69, 37}

\bibitem{Bloch1946}
Bloch F, Hansen W~W and Packard M 1946 {\em Phys. Rev.\/}
  \href{http://dx.doi.org/10.1103/physrev.70.474}{70, 474}

\bibitem{Bloembergen1948}
Bloembergen N, Purcell E~M and Pound R~V 1948 {\em Phys. Rev.\/}
  \href{http://dx.doi.org/10.1103/physrev.73.679}{73, 679}

\bibitem{Bloch1957}
Bloch F 1957 {\em Phys. Rev.\/}
  \href{http://dx.doi.org/10.1103/physrev.105.1206}{105, 1206}

\bibitem{Mollow1969a}
Mollow B~R and Miller M~M 1969 {\em Ann. Phys.\/}
  \href{http://dx.doi.org/10.1016/0003-4916(69)90289-9}{52, 464}

\bibitem{Agarwal1973}
Agarwal G~S 1973 I master equation methods in quantum optics {\em Prog. Opt.\/}
  vol~11 (Elsevier) p~1

\bibitem{Feynman1963}
Feynman R~P and Vernon F~L 1963 {\em Ann. Phys.\/}
  \href{http://dx.doi.org/10.1016/0003-4916(63)90068-x}{24, 118}

\bibitem{Stockburger2002}
Stockburger J~T and Grabert H 2002 {\em Phys. Rev. Lett.\/}
  \href{http://dx.doi.org/10.1103/physrevlett.88.170407}{88, 170407}

\bibitem{Caldeira1981}
Caldeira A~O and Leggett A~J 1981 {\em Phys. Rev. Lett.\/}
  \href{http://dx.doi.org/10.1103/physrevlett.46.211}{46, 211}

\bibitem{Jozsa1994}
Jozsa R 1994 {\em J. Mod. Opt.\/}
  \href{http://dx.doi.org/10.1080/09500349414552171}{41, 2315}

\bibitem{Mukamel1995}
Mukamel S 1995 {\em Principles of nonlinear optical spectroscopy\/} (New York:
  Oxford University Press)

\bibitem{Krantz2019}
Krantz P, Kjaergaard M, Yan F, Orlando T~P, Gustavsson S and Oliver W~D 2019
  {\em Appl. Phys. Rev.\/} \href{http://dx.doi.org/10.1063/1.5089550}{6,
  021318}

\bibitem{Tuorila2009}
Tuorila J and Thuneberg E 2009 {\em J. Phys. Conf. Ser.\/}
  \href{http://dx.doi.org/10.1088/1742-6596/150/2/022092}{150, 022092}

\bibitem{Agarwal1991}
Agarwal G~S, Zhu Y, Gauthier D~J and Mossberg T~W 1991 {\em J. Opt. Soc. Am.
  B\/} \href{http://dx.doi.org/10.1364/josab.8.001163}{8, 1163}

\bibitem{Stockburger2004}
Stockburger J~T 2004 {\em Chem. Phys.\/}
  \href{http://dx.doi.org/10.1016/j.chemphys.2003.09.014}{296, 159}

\bibitem{Stratonovich1957}
{Stratonovich} R~L 1957 {\em Sov. Phys. Dokl.\/} 2, 416

\bibitem{Hubbard1959}
Hubbard J 1959 {\em Phys. Rev. Lett.\/}
  \href{http://dx.doi.org/10.1103/physrevlett.3.77}{3, 77}

\bibitem{Teixeira2019}
Teixeira W~S, Nicacio F and Semi{\~{a}}o F~L 2019 {\em Phys. Rev. A\/}
  \href{http://dx.doi.org/10.1103/physreva.99.032102}{99, 032102}

\end{thebibliography}
	
\end{document}